\documentclass[letterpaper,twocolumn,aps,prl,superscriptaddress,amsmath,amssymb,floatfix]{revtex4-1}
\usepackage{mathptmx}
\usepackage[latin9]{inputenc}
\usepackage{color}
\usepackage{amsmath}
\usepackage{amssymb}
\usepackage{graphicx}
\usepackage{esint}
\usepackage{comment}
\usepackage[unicode=true,
 bookmarks=true,bookmarksnumbered=false,bookmarksopen=false,
 breaklinks=false,pdfborder={0 0 1},backref=false,colorlinks=true]
 {hyperref}
\hypersetup{
 linkcolor=magenta,urlcolor=blue,citecolor=blue,pdfstartview={FitH},hyperfootnotes=false}

\makeatletter

\usepackage{textcomp}
\usepackage{epstopdf}

\pdfpageheight\paperheight
\pdfpagewidth\paperwidth

\@ifundefined{textcolor}{}{%
 \definecolor{BLACK}{gray}{0}
 \definecolor{WHITE}{gray}{1}
 \definecolor{RED}{rgb}{1,0,0}
 \definecolor{GREEN}{rgb}{0,1,0}
 \definecolor{BLUE}{rgb}{0,0,1}
 \definecolor{CYAN}{cmyk}{1,0,0,0}
 \definecolor{MAGENTA}{cmyk}{0,1,0,0}
 \definecolor{YELLOW}{cmyk}{0,0,1,0}
}

\usepackage{xcolor}
\usepackage{soul}
\usepackage{lineno}
\modulolinenumbers[5]

\newcommand{\bra}[1]{\ensuremath{\left\langle#1\right|}}
\newcommand{\ket}[1]{\ensuremath{\left|#1\right\rangle}}

\definecolor{blue}{rgb}{0,0,1}
\definecolor{red}{rgb}{1,0,0}
\definecolor{green}{rgb}{0,1,0}

\usepackage{soul}

\makeatother
\begin{document}
\title{Principles of Optics in the Fock Space: Scalable Manipulation of Giant Quantum States}

\author{Yifang Xu}
\thanks{These authors contributed equally to this work.}
\author{Yilong Zhou}
\thanks{These authors contributed equally to this work.}
\affiliation{Center for Quantum Information, Institute for Interdisciplinary Information Sciences, Tsinghua University, Beijing 100084, China}
\author{Ziyue Hua}
\thanks{These authors contributed equally to this work.}

\author{Lida Sun}
\author{Jie Zhou}
\author{Weiting Wang}
\affiliation{Center for Quantum Information, Institute for Interdisciplinary Information Sciences, Tsinghua University, Beijing 100084, China}

\author{Weizhou Cai}
\affiliation{Laboratory of Quantum Information, University of Science and Technology of China, Hefei 230026, China}

\author{Hongwei Huang}
\author{Lintao Xiao}
\affiliation{Center for Quantum Information, Institute for Interdisciplinary Information Sciences, Tsinghua University, Beijing 100084, China}

\author{Guangming Xue}
    \affiliation{Beijing Academy of Quantum Information Sciences, Beijing, China}
	\affiliation{Hefei National Laboratory, Hefei 230088, China}

\author{Haifeng Yu}
	\affiliation{Beijing Academy of Quantum Information Sciences, Beijing, China}
	\affiliation{Hefei National Laboratory, Hefei 230088, China}

\author{Ming Li}
\email{lmwin@ustc.edu.cn}
\affiliation{Laboratory of Quantum Information, University of Science and Technology of China, Hefei 230026, China}
\affiliation{Hefei National Laboratory, Hefei 230088, China}

\author{Chang-Ling Zou}
\email{clzou321@ustc.edu.cn}
\affiliation{Laboratory of Quantum Information, University of Science and Technology of China, Hefei 230026, China}
\affiliation{Hefei National Laboratory, Hefei 230088, China}

\author{Luyan Sun}
\email{luyansun@tsinghua.edu.cn}
\affiliation{Center for Quantum Information, Institute for Interdisciplinary Information Sciences, Tsinghua University, Beijing 100084, China}
\affiliation{Hefei National Laboratory, Hefei 230088, China}

\begin{abstract}

\textbf{{The manipulation of distinct degrees of freedom of photons plays a critical role in both classical and quantum information processing. While the principles of wave optics provide elegant and scalable control over classical light in spatial and temporal domains, engineering quantum states in Fock space has been largely restricted to few-photon regimes, hindered by the computational and experimental challenges of large Hilbert spaces. Here, we introduce ``Fock-space optics", establishing a conceptual framework of wave propagation in the quantum domain by treating photon number as a synthetic dimension. Using a superconducting microwave resonator, we experimentally demonstrate Fock-space analogues of optical propagation, refraction, lensing, dispersion, and interference with up to 180 photons. These results establish a fundamental correspondence between Schr\"{o}dinger evolution in a single bosonic mode and classical paraxial wave propagation. By mapping intuitive optical concepts onto high-dimensional quantum state engineering, our work opens a path toward scalable control of large-scale quantum systems with thousands of photons and advanced bosonic information processing.}}
\end{abstract}

\maketitle

\noindent The mastery of light, including the generation, manipulation, and detection of light, lies behind vast domains of modern science and enables technologies essential to daily life. The exploration of optical principles, from Newton's geometric analysis of prism dispersion in 1666~\cite{NewtonOpticks} to Young's wave interference demonstration in 1801~\cite{YoungInterference}, laid the foundation for understanding and controlling optical phenomena. Building upon the complementary ray and wave descriptions of light, the unified framework of Maxwell's electromagnetic theory~\cite{Maxwell1865PTRS,Maxwell1873Nature} established the solid ground for engineering optical systems at all scales and led to the prosperity of photonics and information technology we witness today. Despite the complete and mathematically rigorous theoretical foundation provided by Maxwell's equations, optical principles remain indispensable because calculating electromagnetic field distributions at macroscopic scales proves computationally prohibitive and lacks physical intuition. The toolbox of geometric ray tracing and wave interference, therefore, continues to provide efficient, precise, and intuitive design principles for modern optical systems. This powerful intuition and toolbox can be universally extended across diverse physical systems, from water and acoustic waves to de Broglie's matter waves~\cite{DeBroglie}. 

The advent of quantum mechanics unveiled an additional fundamental degree of freedom for light: the quantized energy levels of an electromagnetic mode, which constitute the Fock space, i.e., a Hilbert space spanned by photon-number eigenstates $\ket{n}$. This theoretically infinite-dimensional resource underpins quantum communication~\cite{Azuma2023}, sensing~\cite{LiuXinyuSensing,FockDengXiaowei2024NP100,FockOtherMcCormick2019NatureIontrap},  computation~\cite{CVDVLiu2024arxiv,WeizhouCai2021FRBosonic,BosonicQEC_Joshi_2021}, and  simulation~\cite{FockAppWang2020PRXChemistry,yuan2018synthetic,deng2022observing}, promising quantum-enabled advantages. Harnessing these advantages requires precise control over quantum states in Fock space, which is crucial for approaching the capacity limit in quantum communication and the Heisenberg limit in precision measurements~\cite{QMBraunstein1994Heisenberglimit,QMGiovannetti2004Heisenberglimit,FockDengXiaowei2024NP100}. Previous approaches have relied primarily on bottom-up quantum control strategies, by carefully engineering control Hamiltonians to realize target unitary operations~\cite{UControlKhaneja2005JMRGRAPE,UControlHeeres2017NCGRAPE,UControlEickbusch2022NPECD,UContorlKrastanov2015PRASNAP,UControlHeeres2015PRLSNAP,UControlFosel2020arxivSNAP,UControlKudra2022PRXQSNAP}. This process requires computationally intensive optimization over all relevant energy levels. Although such methods can achieve high precision, the effective manipulation of quantum states in Fock space remains severely constrained by computational resources and experimental complexity. As a result, despite substantial efforts, deterministic control has been limited to a handful of photons~\cite{FockMax2008Nature,FockWang2008PRL,FockMax2009Nature,FockOtherChuYiwen2018NaturePhonon,FockPremaratne2017NCSTIRAP,FockWeitingWang2017PRL,FockAndrew2021ScienceAdvanceFock1,UControlHeeres2015PRLSNAP,UControlKudra2022PRXQSNAP,UControlHeeres2017NCGRAPE,UControlEickbusch2022NPECD,FockBretheau2015SciencePhotonBlockade}, with recent demonstrations of 100-photon Fock states achieved only through non-scalable, probabilistic methods~\cite{FockDengXiaowei2024NP100}. Therefore, there is an urgent need for efficient tools for engineering quantum states in Fock space that can scale to the massive excitations required for practical quantum technologies.

Here, we establish and experimentally demonstrate a framework of Fock-space optics, revealing a profound mathematical duality between quantum dynamics in the photon-number basis and classical optical propagation. Using weak coherent driving of a superconducting cavity that supports up to 180 photons, we show that the quantum evolution equation in the large-photon limit maps directly onto the paraxial wave equation that governs classical beam propagation. This correspondence enables the direct translation of optical principles refined over three centuries to quantum state engineering in Fock space by validating the basic optical tools including propagation, refraction, lensing, dispersion, and interferences. These results demonstrate new features of optics in a synthetic dimension in Fock space and open an intuitive, scalable pathway for manipulating massive quantum excitations, thereby facilitating  the exploration of the advantages in bosonic quantum information processing~\cite{CVDVLiu2024arxiv,WeizhouCai2021FRBosonic,BosonicQEC_Joshi_2021}. 

\bigskip
\noindent\textbf{Principle}

\begin{figure*}
    \centering
    \includegraphics{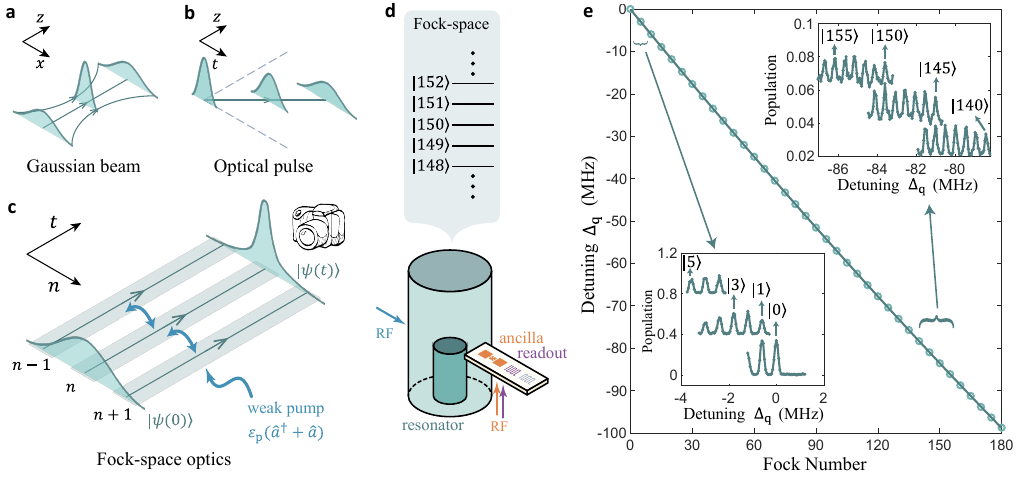}
    \caption{\textbf{Concept and experimental implementation of Fock-space optics.}
    \textbf{a-c}, Illustration of optical propagation in different degrees of freedom: spatial (e.g., \textbf{a,} Gaussian beams), temporal (e.g., \textbf{b,} optical pulses), and the discrete photon-number dimensions (\textbf{c,} Fock-space wavefunctions). 
    \textbf{d}, Schematic of the experimental platform: a superconducting resonator coupled to an ancilla transmon qubit. The resonator encodes quantum states in Fock space, while the ancilla qubit enables both phase manipulation and state-selective readout. 
    \textbf{e}, Calibration of the qubit frequency versus photon number in the resonator, which establishes a Fock-space camera for detecting population along the photon-number axis. Insets: photon-number-splitting peaks of coherent states with different mean photon numbers, $\bar{n}$.
    }
    \label{fig:1}
\end{figure*}

\noindent At its core, Fock-space optics emerges from a deep duality between the quantum dynamics of a single bosonic mode following Schr\"{o}dinger equation and the wave dynamics of optical paraxial propagation. In general, the diffusion equation 
\begin{equation}
        i\frac{\partial}{\partial \zeta}A=D\frac{\partial^2}{\partial \xi^2}A
        \label{eq:diffusion}
\end{equation}
describes the dynamics of a scalar field $A(\xi,\zeta)$ distributed along the direction $\xi$ as it propagates along the axis $\zeta$, with $D$ denoting the diffusion coefficient. For instance, Fig.~\ref{fig:1}a illustrates the propagation of an optical beam, with its dynamics governed by the Helmholtz equation in the paraxial approximation~\cite{kolner1994space}, i.e., Eq.~(\ref{eq:diffusion}) under the substitutions $\zeta=z$, $\xi=x$, and $D=1/2k$ ($k$ is the wave number). This diffusion equation underpins the foundational wave dynamics in optical phenomena, from refraction and diffraction to imaging, which are universal across different platforms or degrees of freedom. As shown in Fig.~\ref{fig:1}b, the same equation can be extended from the spatial dimensions to the temporal domain by interpreting $\xi$ as time~\cite{karpinski2017bandwidth,tirole2023double}. In this context, it describes a temporal pulse in a dispersive medium,  with $\zeta=z$ and $D$ representing the propagation direction and the group velocity dispersion, respectively. 

We now further generalize the wave dynamics to the virtual dimension of the Fock-state space. The quantum state of a single bosonic mode is described by the wavefunction $\ket{\psi} = \sum_n c_n \ket{n}$, which is spanned by the discrete Fock states ${\ket{n}}$ with amplitudes $c_n$. Despite the discrete nature of the Fock space, a correspondence can be drawn to classical discrete optical systems consisting of waveguide arrays~\cite{peschel1998optical, christodoulides2003discretizing}, where photons exhibit discrete diffraction. When a weak pump $\varepsilon_\mathrm{p} (\hat{a}^\dagger + \hat{a})$ is applied to the bosonic mode, with $\hat{a}^\dagger(\hat{a})$ denoting the creation (annihilation) operator of the resonator mode, adjacent Fock states couple together (Fig.~\ref{fig:1}c), which is analogous to nearest-neighbor hopping in discrete waveguide arrays. In the limit of large photon number ($n\gg1$), the amplitudes $c_n$ satisfy a discrete diffusion equation derived from the Schr\"{o}dinger equation [see Sec.~II of the Supplementary Materials for full derivation]:
\begin{eqnarray}
i\frac{\partial}{\partial t} c_{n} & = & \varepsilon_\mathrm{p}\sqrt{{n}}\Delta_{2}[c_{n}]+2\varepsilon_\mathrm{p}\sqrt{{n}} c_{n},
\label{eq:paraxial diffraction}
\end{eqnarray}
where $\Delta_{2}[c_{n}]=c_{n+1}+c_{n-1}-2c_{n}$ is the second-order difference for a discrete variable, corresponding to the second-order differentiation. 

Eliminating the linear drift term via a frame transformation and applying the continuous variable approximation $\Delta_{2}[c_{n}]\approx\partial^2/{\partial n^2}$ for a slowly varying envelope of $c_n$, the formal duality between Eq.~(\ref{eq:paraxial diffraction}) and the diffusion in classical optics [Eq.~(\ref{eq:diffusion})] leads directly to the correspondences:
\begin{equation}
   \xi  \rightarrow n,\,\zeta  \rightarrow  t,\,D  \rightarrow  \sqrt{\bar{n}}\varepsilon_\mathrm{p},  \nonumber
\end{equation}
with $\sqrt{n}\approx \sqrt{\bar{n}}$ for $n\gg1$. This establishes a fundamental analogy in which the Fock state index $n$ plays the role of the spatial dimension $x$, and the quantum state in Fock space maps onto an optical field distribution in optics. Consequently, the probability amplitudes $|c_n|^2$ represent the intensity distribution along the photon-number axis, while the phases $\arg(c_n)$ define the phase profile (or the wavefront).

This duality enables quantum state manipulation in Fock space using the principles of optics. By controlling $\varepsilon_\mathrm{p}=0$, the evolution of the state amplitudes is frozen, halting the propagation in this synthetic dimension. However, we can shape the wavefront by imprinting specific phases on each $c_n$ via tailored interactions, which introduce position-dependent phase shifts analogous to classical optical media/devices. When $\varepsilon_\mathrm{p}\neq0$, the hopping between adjacent Fock states is switched on and the diffusion-like dynamics resumes, allowing the quantum state to propagate and evolve under the influence of the shaped wavefront. Therefore, the combination of phase manipulation and controlled propagation allows us to realize arbitrary photon-number distributions, subject only to the fundamental diffraction limit inherent to wave optics.

\bigskip
\noindent\textbf{Experimental setup}

\noindent Building on this duality, we experimentally implement Fock-space optics in a superconducting circuit quantum electrodynamics platform~\cite{CQEDWallraff2004Nature,CQEDBlais2021RMP}. As depicted in Fig.~\ref{fig:1}d, a high-coherence microwave resonator, with frequency $\omega_a/2\pi = 6.140\,\mathrm{GHz}$ and coherence time $T_1 = 1.6\,\mathrm{ms}$, serves as the bosonic mode for encoding the photon-number wavefunction. A weak microwave pump, detuned by $\Delta$ with a strength of $\varepsilon_\mathrm{p}/2\pi = 0.88\,\mathrm{MHz}$, induces coherent propagation of the wavefunction through inter-Fock-state couplings. A dispersively coupled transmon qubit~\cite{TransmonKoch2007PRA} provides the tunable nonlinearity essential for both wavefront shaping and imaging in the Fock space. The corresponding system Hamiltonian is ($\hbar=1$) 
\begin{equation}\begin{split}\label{eq:Hamitonian}
    \hat{H} & = 
     \Delta \hat{a}^\dagger\hat{a} 
     - \frac{K_\mathrm{4}}{2}\hat{a}^{\dagger2}\hat{a}^2
      \\
    & -\chi \hat{a}^\dagger\hat{a} \ket{\mathrm{e}} \bra{\mathrm{e}} 
    +\frac{K_\mathrm{e}}{2}\hat{a}^{\dagger2}\hat{a}^2 \ket{\mathrm{e}} \bra{\mathrm{e}}  \\
    & +\varepsilon_\mathrm{p}(\hat{a}^\dagger+\hat{a}),    
\end{split}\end{equation}
where $\ket{\mathrm{e}}$ denotes the qubit's excited state, $K_\mathrm{4}/2\pi=2.18\,\mathrm{kHz}$ is the resonator's self-Kerr coefficient, $\chi/2\pi=0.596\,\mathrm{MHz}$ and $K_\mathrm{e}/2\pi=0.52\,\mathrm{kHz}$ represent the cross-Kerr terms between the resonator and the qubit.

In this Fock-space optical framework, the terms in the first and second lines of Eq.~(\ref{eq:Hamitonian}) emulate an optical medium that generates a photon number $a^{\dagger}a$-dependent phase accumulation: $\Delta$ yields a linear refractive index gradient along the $n$-axis (Fig.~\ref{fig:1}c), while $K_{4}$ provides a quadratic refractive index distribution for lens-like effects. The terms in the second line enable a switchable refractive index distribution, whereas the last term facilitates switchable diffusion. Additionally, the terms in the second line also indicate a nonlinear coupling between the Fock-space wavefunction and the qubit, allowing a photon-number-resolved readout of the wavefunction~\cite{NumSplSchuster2007Nature}, i.e., the detection of $\ket{\psi}$, via selective population measurements through frequency-tuned qubit pulses (see Sec.~IV of the Supplementary Materials for experimental setup details). As shown in Fig.~\ref{fig:1}e, the successful detection of Fock states up to $n=180$ establishes an effective Fock-space camera. 

Leveraging this experimental platform and the clear analogies to classical optics, we now examine the principles of optics in Fock space by demonstrating the most essential wave phenomena and devices, beginning with basic optical elements.

\begin{figure}[t]
    \centering
    \includegraphics{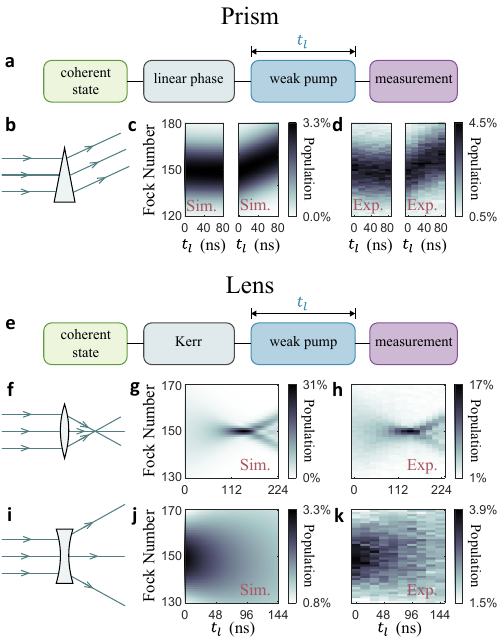}
    \caption{\textbf{Prism and lens elements in Fock space.} 
    \textbf{a,} Sequence and \textbf{b,} concept of the prism. 
    A coherent state (e.g., $\bar{n}=150$) is prepared via a strong displacement, followed by a weak single-photon pump. A linear phase accumulation is imprinted on each Fock state, controlled by either the phase of the weak pump or a period of detuned free evolution. Finally, the population distribution across the Fock states is measured.
    \textbf{c,} Simulation and \textbf{d,} experimental results for the Fock-space prism.
    \textbf{e,} Sequence and \textbf{f (i),} concept of the convex (concave) lens. 
    After preparing a coherent state (e.g., $\bar{n}=150$), the resonator evolves freely under the self-Kerr effect, followed by a weak single-photon pump. During the intermediate free evolution period, the combined effects of self-Kerr nonlinearity and detuning lead to a quadratic phase accumulation centered around $n=150$, analogous to a classical lens.
    By inverting the phase of the pump, the process is transformed to a concave lens.
    \textbf{g (j),} Simulations and \textbf{h (k),} experiments demonstrating the focusing (diverging) effect on quantum states.
    Using the highly focused state from the convex lens, Fock state $\ket{150}$ can be prepared with a success rate up to 17\%.
    }
    \label{fig:2}
\end{figure}

\bigskip
\noindent\textbf{Basic optical elements in Fock space}

\noindent A fundamental concept in classical optics is the Gaussian beam, which approximates paraxial wave propagation with a Gaussian intensity profile transverse to the direction of travel. This enables the practical realization of optical phenomena such as diffraction, refraction, and focusing. In Fock space, a coherent state with $c_n = \frac{\alpha^n}{\sqrt{n!}} e^{-|\alpha|^2/2}$ (where $\alpha = \sqrt{\bar{n}}$) serves as an effective analog to a Gaussian beam when the mean photon number $\bar{n}$ is large. In this limit, the Poissonian photon-number distribution approaches a Gaussian distribution. As illustrated in Fig.~\ref{fig:2}a, we demonstrate the propagation and refraction of a quantum state in Fock space by tuning the pump field. An initial coherent state $\ket{\psi}$ with $\bar{n}=150$, resembling a Gaussian beam input, is prepared by applying a displacement operation to the vacuum. A weak single-photon pump is then applied to evolve the state, during which the population distribution across Fock states is measured using the ancilla qubit. We should note that the pump amplitude $|\varepsilon_\mathrm{p}|$ and the interaction duration $t_l$ determine the diffusion, while the phase of the drive $\phi_\mathrm{p}=\mathrm{arg}(\varepsilon_\mathrm{p})$ induces an effective gradient phase on the input state via the gauge transformation $c_n\rightarrow c_ne^{\mathrm{i}n\phi_\mathrm{p}}$. This imprints a linear phase $\varphi(n) = n\phi_\mathrm{p}$, which is directly equivalent to the spatially-dependent phase imposed by a prism in optics, as illustrated in Fig.~\ref{fig:2}b.

As shown in Figs.~\ref{fig:2}c and \ref{fig:2}d, the temporal evolution under this weak pump exhibits straight-line propagation analogous to a Gaussian beam in free space, showing excellent agreement between theoretical and experimental results. When $\phi_\mathrm{p}=0$, the state evolves with its profile and center location (mean photon number) fixed, as expected for a freely propagating Gaussian beam in optics. In contrast, refraction of the beam is observed for $\phi_\mathrm{p}=\pi/2$, as the center location increases linearly with $|\varepsilon_\mathrm{p}| t_l$. These behaviors can be well explained as the weak pump induces a displacement of the coherent state, changing its amplitude from $\alpha=\sqrt{\bar{n}}$ to $\sqrt{\bar{n}}+\mathrm{i}\varepsilon_\mathrm{p} t_l$. Consequently, the mean photon number changes to $\bar{n}-2\sqrt{\bar{n}}\mathrm{sin}\phi_\mathrm{p}\varepsilon_\mathrm{p} t_l$ by neglecting the second-order terms $\mathcal{O}{|\varepsilon_\mathrm{p} t_l|^2}$ in the weak pump limit. 

\begin{figure}[t]
    \centering
    \includegraphics{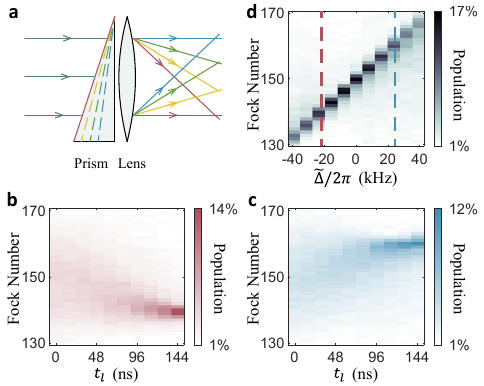}
    \caption{\textbf{{Newton's prism experiment in Fock space.}}  
    \textbf{a,} Concept of Newton's prism experiment. By combining a prism and a lens, light of different colors is deflected and focuses onto different points on the focal plane.
    In Fock space, the microwave frequency $\omega_\mathrm{p}$ of the single-photon pump, corresponding to different $\Delta$ in Eq.~(\ref{eq:Hamitonian}), acts as the ``color", determining the deflection angle of the prism.
    \textbf{b, c,} Experimental results demonstrating the focusing of {a coherent state ($\bar{n}=150$)} at different Fock numbers for different $\tilde\Delta$. For $\tilde\Delta/2\pi=-23\,\mathrm{kHz}$ (\textbf{b}) and $23\,\mathrm{kHz}$ (\textbf{c}), the quantum state is focused to $n=140$ and $n=160$, respectively, at time $t_l=144\,\mathrm{ns}$.
    \textbf{d,} Fock state populations at the focal plane for different detunings. The red and blue dashed lines correspond to the cases in \textbf{b} and \textbf{c}, respectively.
    }
    \label{fig:3}
\end{figure}

The lens constitutes another fundamental optical element that imparts a quadratic phase shift to an incident beam, transforming its wavefront to spherical curvature and enabling focusing or defocusing under the paraxial approximation. An ideal lens operation in Fock space should provide a quadratic phase profile to the quantum state via the transformation $c_n\rightarrow c_ne^{\mathrm{i}(n-\bar{n})^2\phi_0}$, where the quadratic phase coefficient $\phi_0$ determines the focal length (Figs.~\ref{fig:2}f and \ref{fig:2}i). We experimentally realize this transformation using a coherent state input with $\bar{n}=150$ by sequentially applying phase accumulation and controlled propagation, leveraging the resonator's self-Kerr nonlinearity in combination with pump detuning, as depicted in Fig.~\ref{fig:2}e. A quadratic phase $\varphi(n)=-\frac{K_4}{2}t_\varphi n^2+\Delta t_\varphi n$ is accumulated after a free evolution period of duration $t_\varphi = 4684$~ns, where the detuning $\Delta/2\pi=\Delta_\mathrm{L} / 2\pi = 0.33$~MHz is precisely chosen to center the quadratic phase at $n=150$, matching the input state. A subsequent application of a weak single-photon pump then induces propagation dynamics that reveal the lensing behavior. As shown in Figs.~\ref{fig:2}g and \ref{fig:2}h, the temporal evolution under this weak pump, configured as a convex lens, exhibits tight focusing of the quantum state into a narrow range of Fock numbers. The experimental data align closely with the theoretical simulations. Although deviation arises due to dissipation and measurement errors (see Sec.~V.C of the Supplementary Materials), the lensing of the coherent state enables efficient preparation of high-photon-number Fock states. For instance, we achieve a success rate of $17\%$ for generating Fock state $|150\rangle$, significantly exceeding previous methods (e.g., 2.2\% for Fock state $\ket{100}$ in Ref.~\cite{FockDengXiaowei2024NP100}).

A profound insight into optics in Fock space emerges when the pump phase is set to $\phi_\mathrm{p}=\pi$. Due to the time-reversal symmetry of the Schr\"{o}dinger equation, the inverted sign of the diffusion interaction leads to the forward propagation of the conjugated quantum state $\bra{\psi}$. Equivalently, the phase $\phi_\mathrm{p}=\pi$ implements a time-reversed evolution (backward propagation) for the state $\ket{\psi}$, converting convergence into divergence, as shown in Fig.~\ref{fig:2}i. As observed in Figs.~\ref{fig:2}j and \ref{fig:2}k, the resulting concave lens configuration spreads the quantum state to a broader range of photon numbers. This demonstration reveals a unique and elegant mechanism for testing the time-reversed optics in Fock space, a feature with no direct classical counterpart.

\begin{figure*}
    \centering
    \includegraphics{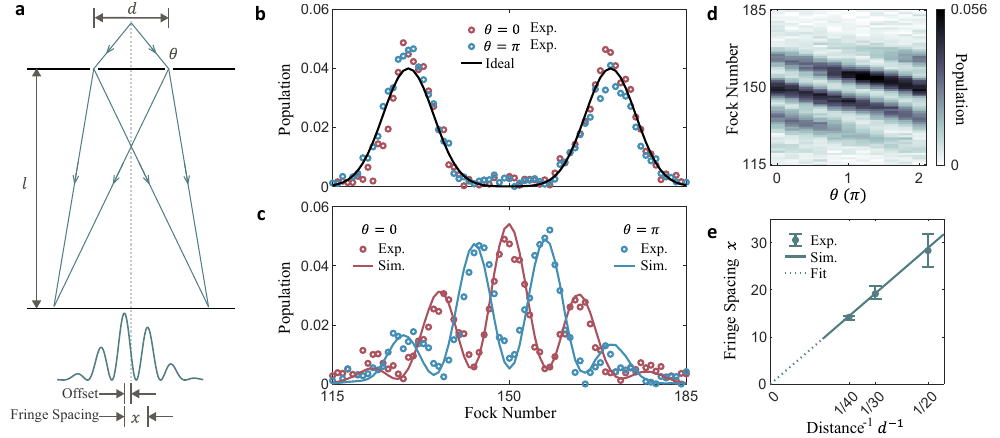}
    \caption{\textbf{Young's double-slit interference in Fock space.} \textbf{a,} Concept of Young's double-slit interference experiment. Coherent light diffracts from two slits and creates interference fringes in the overlapping region on a distant screen. The fringe spacing is inversely proportional to the slit separation $d$ and the pattern shifts with the relative phase $\theta$ between the two slits.
    \textbf{b,} Ideal and experimental results of the double-Gaussian (DG) state $\ket{\psi_\mathrm{DG}}$, analogous to the double slits. The state consists of two Gaussian peaks of equal height centered at $n_1=130$ and $n_2=170$, shown for $\theta=0$ and $\pi$.
    \textbf{c,} Simulated and experimental interference fringes at time $t_l=1.1\,\mathrm{\mu s}$ for $\theta=0$ and $\theta=\pi$.
    \textbf{d,} Experimental interference fringes for different phases $\theta$ between the slits. The change of phase results in an overall shift in the fringe pattern with a $2\pi$ period, causing the central Fock state to transition from coherent enhancement to subtraction and back again.
    \textbf{e,} Fringe spacing as a function of the distance between the two peaks of the DG state (analogous to the slit separation). The spacings, extracted by fitting the data to a Gaussian-envelope cosine function, confirm the inverse proportionality. Experimental data and fits are provided in Fig.~S8 of the Supplementary Materials.
    }
    \label{fig:4}
\end{figure*}

\begin{figure}[t]
    \centering
    \includegraphics{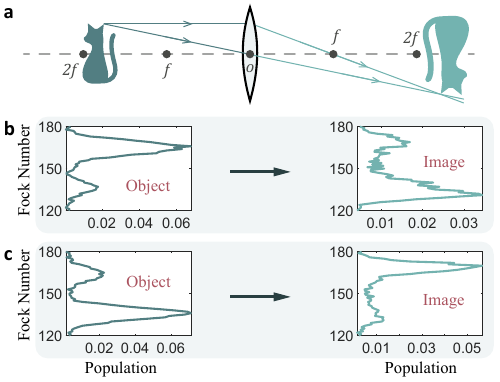}
    \caption{\textbf{Imaging in Fock space.} 
    \textbf{a,} Concept of imaging in classical optics for the case $f<u<2f$, where $u$ is the object distance and $f$ is the focal length.
    \textbf{b,} Left: Fock-space distribution of the ``object" state $\ket{\psi_\mathrm{DG}}$, featuring two Gaussian peaks (standard deviation $\sigma=5$) centered at $n_1=135$ and  $n_2=165$ with a peak height ratio of $1:4$.
    Right: Fock-space distribution of the inverted, magnified, real ``image" state observed at the image distance predicted by Eq.~(\ref{eq:image}). 
    The separation between the peak centers confirms the magnification factor $t_v/t_u$.
    \textbf{c,} ``Object" and ``image" states from a similar experiment using a DG state with a peak height ratio of $4:1$.
    The cosine similarity between the ideal and measured photon number distributions of the image is 84\% (\textbf{b}) and 90\% (\textbf{c}), respectively.
    }
    \label{fig:5}
\end{figure}

\bigskip
\noindent\textbf{Newton's prism experiment in Fock space}

\noindent Building upon the basic optical elements in Fock space, we explore the prism-lens combination system in Fig.~\ref{fig:3} to reproduce the elegant Newton's prism experiments~\cite{NewtonOpticks}. As shown in Fig.~\ref{fig:3}a, Newton demonstrated that a prism disperses light spatially based on its wavelength. 
This principle of spectral-to-spatial conversion, achieved through wavelength-dependent phase accumulation, is now widely applied in spectroscopic analysis.

Here, we realize a similar spectral-to-photon-number conversion by varying the pump frequency $\omega_\mathrm{p}$. By combining the sequences from Fig.~\ref{fig:2} and leveraging the fact that the Fock-space prism and lens commute, this single parameter $\omega_\mathrm{p}$ simultaneously controls two key operations: a focal axis alignment ($\Delta_\mathrm{L}$) for the lens and a linear phase accumulation ($\tilde\Delta=\Delta-\Delta_\mathrm{L}$) that acts as the prism. This linear phase $\tilde\Delta$ is analogous to the wavelength-dependent refractive index in Newton's prism.
The frequency-dependent dispersion in Fock space arises because different values of $\tilde\Delta$ impart a distinct linear phase gradient to the quantum state, thereby encoding frequency information into the Fock-space wavefront. Consequently, the subsequent lens focuses the quantum state with $\bar{n}$ photons to a specific photon number of $n\approx \bar{n}+\tilde\Delta/K_4$. This creates a one-to-one map between the pump frequency and the output photon number, realizing spectroscopic dispersion entirely within the synthetic dimension.

We demonstrate this dispersion by examining the evolution of a coherent state ($\bar{n}=150$) for two specific cases in Figs.~\ref{fig:3}b and \ref{fig:3}c. For a phase accumulation of $\tilde\Delta/2\pi=-23\,\mathrm{kHz}$, the quantum state focuses at $n=140$, while $\tilde\Delta/2\pi=23\,\mathrm{kHz}$ shifts the focus to $n=160$. The linear relationship between the detuning and the focal position, predicted by the principles of optics, is confirmed through systematic measurements, as summarized in Fig.~\ref{fig:3}d.

\bigskip
\noindent\textbf{Fock-space double-slit interference}

\noindent Young's double-slit experiment~\cite{YoungInterference} of 1801 provided decisive evidence for the wave nature of light and later became a conceptual cornerstone of quantum mechanics through demonstrations of wave-particle duality. As shown in Fig.~\ref{fig:4}a, the optical setup consists of two coherent point sources that generate interference fringes on the observation screen, where the visibility (modulation depth) and location of the fringes manifest the optical coherence and the wave nature. To demonstrate Young's double-slit interference entirely within Fock space, we need an input quantum state that mimics two spatially separated narrow beams. This corresponds to the superposition of two narrow peaks along the $n$-axis in Fock space. We create such a double-Gaussian (DG) state $|\psi_{\text{DG}}\rangle$ with two narrow peaks separated by $d$ Fock numbers:
\begin{equation}
|\psi_{\text{DG}}\rangle = \frac{1}{\sqrt{2}}\left(|G_{n_1}\rangle + e^{i\theta}|G_{n_2}\rangle\right),
\end{equation}
where $|G_{n_{1(2)}}\rangle$ represents a Gaussian distribution centered at photon number $n_{1(2)}$ with a standard deviation of $\sigma = 5$. 

Figure~\ref{fig:4}b presents the distribution of the state $|\psi_{\text{DG}}\rangle$ with peaks at $n_1 = 130$ and $n_2 = 170$ (separation $d = 40$), prepared via  the phase-space slingshot approach (see Sec.~III of the Supplementary Materials). When applying the weak pump immediately after state preparation, each Gaussian component undergoes diffraction according to the evolution equation [Eq.~(\ref{eq:diffusion})]. The narrow width $\sigma = 5$ of the two components ensures substantial diffraction, causing them to overlap and interfere in the intermediate region. The resulting periodic modulation of the population distribution, as shown in Fig.~\ref{fig:4}c, depends critically on the relative phase $\theta$ between the two components. As in optics, a well-defined $\theta$ is necessary for interference; a mixed state with an uncertain phase shows no fringes (Fig.~S7c of the Supplementary Materials). For $\theta = 0$, constructive interference enhances the population at specific photon numbers (e.g., $n=150$), while $\theta = \pi$ results in destructive interference and population suppression. The fringe pattern shifts continuously with $\theta$ over a $2\pi$ range (Fig.~\ref{fig:4}d), providing direct evidence of coherent superposition between the two Gaussian components. The fringe spacing is given by $x \propto \varepsilon_\mathrm{p} t_l/d$, analogous to classical optics. In Fig.~\ref{fig:4}e, the fringe spacings are extracted from Gaussian-enveloped cosine fits under different conditions of $d$, agreeing well with both analytical and numerical results.

\bigskip
\noindent{\textbf{Fock-space imaging}}

\noindent Having established the basic optical elements in Fock space, we now demonstrate their combined use for quantum state manipulations that are intractable with conventional approaches. In Fock space, we realize the imaging through a lens sandwiched between two free propagation segments, as illustrated in Fig.~\ref{fig:5}a.
The propagation times, $t_u$ and $t_v$, correspond to the object and image distances in a classical imaging system and satisfy the standard imaging condition:
\begin{equation}\label{eq:image}
     \frac{1}{t_u} + \frac{1}{t_v} = \frac{1}{t_f},
\end{equation}
where $t_f=144\,\mathrm{ns}$ is the focal duration time of our Kerr-nonlinearity lens, as calibrated in Fig.~\ref{fig:2}h. Through the imaging principle, we can transform any quantum state with a Fock-space distribution $c_n$ into a magnified and inverted image $c^{\prime}_n=c_{-M(n-n_0)+n_0}$, with $n_0=150$ denoting the center position of the lens and $M={t_v}/{t_u}$ being the magnification factor. This imaging system exemplifies the power of the optical framework: it performs a unitary transformation across a large Hilbert space which coherently inverts and magnifies arbitrary quantum state distributions, a task that would require optimizing thousands of parameters using traditional quantum control methods.

As examples, we image quantum states with DG distributions $\ket{\psi_\mathrm{DG}}$. The object state (left panels of Figs.~\ref{fig:5}b and \ref{fig:5}c) consist of two Gaussian peaks centered at $n_1=135$ and $n_2=165$ ($d=30$), each with a standard deviation of $\sigma=5$. The peak height ratios are $1:4$ and $4:1$ for Figs.~\ref{fig:5}b and \ref{fig:5}c, respectively. After propagating with an object time $t_u=250\,\mathrm{ns}$ and an image time $t_v = 1/(1/t_f-1/t_u)=340\,\mathrm{ns}$, we obtain the resulting image states (right panels of Figs.~\ref{fig:5}b and \ref{fig:5}c). As expected, the imaging system produces an inverted and magnified image with a larger distance. The peak separations increase to 
$d'=38.2$ and $37.7$ for Figs.~\ref{fig:5}b and \ref{fig:5}c, respectively, corresponding to magnification factors of 
$1.27$ and $1.26$ that are in good agreement with the expected value of $M=t_v/t_u=1.36$. To quantify the imaging quality in Fock space, we use the cosine similarity, defined as $\frac{\textbf{A}\cdot\textbf{B}}{|\textbf{A}||\textbf{B}|}$, where $\textbf{A}$ and $\textbf{B}$ are vectors representing the ideal and measured photon number distributions of the image, respectively. The ideal distribution $\textbf{A}$ is derived from the imaging model in Eq.~(\ref{eq:image}), with the positions of the two Gaussian centers and the scaled standard deviation $M\sigma$ given by the theory. The high similarity values of 84\% and 90\% for Figs.~\ref{fig:5}b and \ref{fig:5}c, respectively, demonstrate the high-fidelity nature of the imaging.

\bigskip
\noindent\textbf{Discussions}

\noindent Our demonstration of optical principles in Fock space establishes a fundamental duality between quantum state evolution in a single bosonic mode and classical paraxial wave propagation. Through experimental realization of propagation, refraction, lensing, dispersion, and interference phenomena with up to 180 photons, we show that quantum dynamics in the Fock space can be understood and controlled using optical principles. This Fock optics framework also highlights a unique feature of time reversal propagation due to the conjugate symmetry of quantum mechanics, promising novel phenomena in such a synthetic dimension. This duality transforms quantum state engineering in high-dimensional Hilbert spaces from a computational challenge into an intuitive design process. Guided by the intuitive ray or wave optical design, we can manipulate quantum states by combining and arranging basic elements in the optical toolbox, without solving the full Schr\"{o}dinger equation. This approach establishes a scalable framework for efficient state engineering, reaching an unprecedented regime of thousands or even more photon numbers~\cite{Li2026arXiv2601.05118}. This is evidenced by our preparation of 150-photon Fock states with a 17\% success rate, representing an order of magnitude improvement over existing methods. The resulting highly squeezed states constitute a powerful and scalable resource for quantum-enhanced sensing [Sec.~VII of the Supplementary Materials].

Looking forward, the Fock-space optics toolbox can be expanded to include elements such as mirrors, gratings, and metasurfaces~\cite{chen2016review}, promising even more powerful control over quantum photon states. Such compound optical systems could realize protocols currently intractable for conventional quantum control approaches, such as the operations that amplify states encoded with low photon numbers or perform complex quantum state transformations. Furthermore, the framework is readily applicable to other high-dimensional quantum systems, including mechanical oscillators~\cite{FockOtherChuYiwen2018NaturePhonon}, cold atoms~\cite{Chen2015}, and trapped ions~\cite{FockOtherMcCormick2019NatureIontrap}, and it can be naturally extended beyond single-mode cavities. In these diverse regimes, Fock-space optics offers a powerful approach for manipulating intriguing quantum many-body states.

\smallskip{}


%

\bigskip{}

\noindent \textbf{\large{}{}Data availability}{\large\par}

\noindent All data generated or analyzed during this study are available within the paper and its Supplementary Materials. Further source data will be made available on reasonable request.

\smallskip{}

\noindent \textbf{\large{}{}Code availability}{\large\par}

\noindent The code used in this study is available from the corresponding author upon reasonable request.

\smallskip{}

\bigskip
\noindent\textbf{Acknowledgements}
This work was funded by the National Natural Science Foundation of China 
(Grants No.~92165209, 92265210, 92365301, 92365206, 12204052, 12474498, 92565301, 12550006, 11925404),  Innovation Program for Quantum Science and Technology (Grant No.~2021ZD0300200, 2021ZD0301800, and 2024ZD0301500). This work was also supported by the Fundamental Research Funds for the Central Universities and USTC Research Funds of the Double First-Class Initiative. This work was partially carried out at the USTC Center for Micro and Nanoscale Research and Fabrication, and  numerical calculations were performed at the Supercomputing Center of USTC.

\bigskip
\noindent\textbf{Author contributions}
C.-L.Z. and M.L. conceived the experiment and provided theoretical support. Y.X. and Y.Z. performed the experiment, analyzed the data, and carried out the numerical simulations under the supervision of Luyan Sun. Z.H. provided simulation support. Y.Z. developed the FPGA technique. Li.S. helped to calibrate the system.
J.Z., W.W., W.C., H.H., L.X contributed to the experimental support.
W.C. fabricated the 3D cavity. G.X. and H.Y. fabricated the tantalum transmon qubits. Y.X., Y.Z., Z.H., M.L., C.-L.Z. and Luyan Sun wrote the paper with input from all authors. C.-L.Z. and Luyan Sun supervised the project.

\smallskip{}

\noindent \textbf{\large{}{}Competing interests}{\large\par}

\noindent The authors declare no competing interests.

\smallskip{}

\noindent \textbf{\large{}{}Additional information}{\large\par}

\noindent \textbf{Supplementary Materials} The online version contains Supplementary Materials.

\noindent \textbf{Correspondence and requests for materials} should be addressed to M.L., C.-L.Z., or Luyan Sun.

\end{document}


\title{Supplementary Materials to: ``Principles of Optics in the Fock Space: Scalable Manipulation of Giant Quantum States"}

\author{Yifang Xu}
\thanks{These authors contributed equally to this work.}
\author{Yilong Zhou}
\thanks{These authors contributed equally to this work.}
\affiliation{Center for Quantum Information, Institute for Interdisciplinary Information Sciences, Tsinghua University, Beijing 100084, China}
\author{Ziyue Hua}
\thanks{These authors contributed equally to this work.}

\author{Lida Sun}
\author{Jie Zhou}
\author{Weiting Wang}
\affiliation{Center for Quantum Information, Institute for Interdisciplinary Information Sciences, Tsinghua University, Beijing 100084, China}

\author{Weizhou Cai}
\affiliation{CAS Key Laboratory of Quantum Information, University of Science and Technology of China, Hefei 230026, China}

\author{Hongwei Huang}
\author{Lintao Xiao}
\affiliation{Center for Quantum Information, Institute for Interdisciplinary Information Sciences, Tsinghua University, Beijing 100084, China}

\author{Guangming Xue}
    \affiliation{Beijing Academy of Quantum Information Sciences, Beijing, China}
	\affiliation{Hefei National Laboratory, Hefei 230088, China}

\author{Haifeng Yu}
	\affiliation{Beijing Academy of Quantum Information Sciences, Beijing, China}
	\affiliation{Hefei National Laboratory, Hefei 230088, China}

\author{Ming Li}
\email{lmwin@ustc.edu.cn}
\affiliation{CAS Key Laboratory of Quantum Information, University of Science and Technology of China, Hefei 230026, China}
\affiliation{Hefei National Laboratory, Hefei 230088, China}

\author{Chang-Ling Zou}
\email{clzou321@ustc.edu.cn}
\affiliation{CAS Key Laboratory of Quantum Information, University of Science and Technology of China, Hefei 230026, China}
\affiliation{Hefei National Laboratory, Hefei 230088, China}

\author{Luyan Sun}
\email{luyansun@tsinghua.edu.cn}
\affiliation{Center for Quantum Information, Institute for Interdisciplinary Information Sciences, Tsinghua University, Beijing 100084, China}
\affiliation{Hefei National Laboratory, Hefei 230088, China}

\maketitle

\tableofcontents

\newpage

\renewcommand{\thefigure}{S\arabic{figure}}
\setcounter{figure}{0}
\renewcommand{\thetable}{S\arabic{table}}
\setcounter{table}{0}  

\section{System Hamiltonian}
 
Our experimental device has a structure similar to those reported in Refs.~\cite{JieZhou2024PRLQuantumStateTransfer,ChenZJ2025SciAdvOpenGRAPE}, consisting of three high-coherence 3D aluminum cavities and four transmon qubits~\cite{TransmonKoch2007PRA}, each qubit with its own readout resonator. In this work, we employ the components of one 3D cavity and one transmon, dispersively coupled to act as the ``resonator" and the ``ancilla" depicted in Fig.~1d in the main text, respectively. 
The ancilla's readout resonator is used exclusively for qubit readout and will not be included in the theoretical model described below.

The Hamiltonian of the cavity-qubit system is expressed as:
\begin{equation}\label{eq:Ham:all}
    \hat{H}_0 = \hat{H}_\mathrm{a0} + \hat{H}_\mathrm{q0} + \hat{H}_\mathrm{i} + \hat{H}_\mathrm{d},
\end{equation}
where $\hat{H}_\mathrm{a0}$, $\hat{H}_\mathrm{q0}$, $\hat{H}_\mathrm{i}$, and $\hat{H}_\mathrm{d}$ represent the Hamiltonians of the cavity, the transmon qubit, the interaction, and the drive on the cavity, respectively. 
Explicitly ($\hbar = 1$):
\begin{equation}\label{eq:Ham:Hc}
    \hat{H}_\mathrm{a0}=\omega_\mathrm{a0}\hat{a}^\dagger\hat{a},
\end{equation}
\begin{equation}\label{eq:Ham:Hq}
    \hat{H}_\mathrm{q0}=\omega_\mathrm{q0}\hat{b}^\dagger\hat{b}-\frac{K_\mathrm{q4}}{2}\hat{b}^{\dagger2}\hat{b}^2 + \frac{K_\mathrm{q6}}{6}\hat{b}^{\dagger3}\hat{b}^3,
\end{equation}
\begin{equation}\label{eq:Ham:Hi}
    \hat{H}_\mathrm{i} = g(\hat{a}^\dagger\hat{b}+ \hat{b}^\dagger\hat{a}),
\end{equation}
\begin{equation}\label{eq:Ham:Hd}
    \hat{H}_\mathrm{d} = \varepsilon_\mathrm{d}\cos(\omega_\mathrm{d}t)(\hat{a}^\dagger+\hat{a}).
\end{equation}
Here, $\hat{a}^\dagger(\hat{a})$ and $\hat{b}^\dagger(\hat{b})$ are the creation (annihilation) operators of the cavity and the qubit, respectively; $\omega_\mathrm{a0}$ and $\omega_\mathrm{q0}$ are the frequencies of the cavity and the qubit, respectively; $K_\mathrm{q4}$ and $K_\mathrm{q6}$ are the 4th-order and 6th-order Kerr nonlinearities of the transmon, resulting from its cosine potential; $g$ is the coupling strength; $\varepsilon_\mathrm{d}$ is the drive strength; and $\omega_\mathrm{d}$ is the drive frequency.

Following the Bogoliubov approach to handling the Jaynes-Cummings Hamiltonian \cite{CQEDBlais2021RMP}, we perform the transformation $\hat{a}\rightarrow\hat{a}+(g/\Delta_0)\hat{b}$ and $\hat{b}\rightarrow\hat{b}-(g/\Delta_0)\hat{a}$, where $\Delta_0 = \omega_\mathrm{a0} - \omega_\mathrm{q0}$. Neglecting the fast-rotating terms, the system Hamiltonian becomes: 
\begin{equation}
    \hat{H}^\prime \approx \omega_\mathrm{a}\hat{a}^\dagger\hat{a} 
    - \frac{K_\mathrm{4}}{2}\hat{a}^{\dagger2}\hat{a}^2 
    - \frac{K_\mathrm{6}}{6}\hat{a}^{\dagger3}\hat{a}^3
    + \omega_\mathrm{q}\hat{b}^\dagger\hat{b}
    - \frac{K_\mathrm{q4}}{2}\hat{b}^{\dagger2}\hat{b}^2 
    + \frac{K_\mathrm{q6}}{6}\hat{b}^{\dagger3}\hat{b}^3
    - \chi\hat{a}^\dagger\hat{a}\hat{b}^\dagger\hat{b}
    +\frac{K_\mathrm{e}}{2}\hat{a}^{\dagger2}\hat{a}^2\hat{b}^\dagger\hat{b} 
    +\chi_\mathrm{f}\hat{a}^{\dagger}\hat{a}\hat{b}^{\dagger2}\hat{b}^2 
    +\hat{H}_\mathrm{d}+\cdots.
\end{equation}
The parameters in the Hamiltonian are calibrated experimentally and given in Table~\ref{tab:allparameters}.
It is worth noting that although the 6th-order coefficient $ K_\mathrm{q6} $ for the qubit is positive, the induced 6th-order coefficient $ K_6 $ in the cavity is negative in both the experiment and the numerical simulation (see Sec.~\ref{sec:selfKerr}). 
This can be attributed to the synthesis of fast-rotating higher-order processes~\cite{HPhotonTimo2022PRAppliedHighOrderKerr}. For instance, combining two 4th-order processes can give rise to a 6th-order effect.

By reducing the transmon to a two-level system 
$\{\ket{\mathrm{g}},~\ket{\mathrm{e}}\}$ and moving into the rotating frame defined by 
$\omega_\mathrm{d}$ and $\omega_\mathrm{q}$, we obtain the Hamiltonian:
\begin{equation}\begin{split}\label{eq:Hamitonian}
    \hat{H} & = 
     \Delta \hat{a}^\dagger\hat{a} 
     - \frac{K_\mathrm{4}}{2}\hat{a}^{\dagger2}\hat{a}^2
     - \frac{K_\mathrm{6}}{6}\hat{a}^{\dagger3}\hat{a}^3 
     -\chi \hat{a}^\dagger\hat{a} \ket{\mathrm{e}} \bra{\mathrm{e}} 
    +\frac{K_\mathrm{e}}{2}\hat{a}^{\dagger2}\hat{a}^2 \ket{\mathrm{e}} \bra{\mathrm{e}}  
     +\epsilon_\mathrm{d}(\hat{a}^\dagger+\hat{a}).    
\end{split}\end{equation}
Here, $K_4$ is obtained by performing Ramsey experiments on the cavity, i.e., first preparing $ ( \ket{0}+\ket{N} ) / \sqrt{2} $ states for small Fock numbers $N$ via the GRAPE method~\cite{UControlHeeres2017NCGRAPE,UControlKhaneja2005JMRGRAPE}, then fitting the frequency differences for various $N$. 
$K_6$ is calibrated using a Fock-space Newton's prism experiment, as detailed in Sec.~\ref{sec:selfKerr}. $\chi$ and $K_\mathrm{e}$ are calibrated by fitting the results of the number-splitting experiments, spanning from Fock state $\ket{0}$ to $\ket{180}$ (see Sec.~\ref{sec:camera}).

\section{Wave equation in the Fock space}
\subsection{Paraxial diffraction of the optical field}

The behavior of a monochromatic optical field is governed by the Helmholtz equation:
\begin{eqnarray}
\left(\nabla^{2}+k^{2}\right)\mathbf{E} & = & 0.
\label{eq:Helmholtz}
\end{eqnarray}
Assuming that the field propagates along the $z$ direction, the electric field has the form of:
\begin{eqnarray}
\mathbf{E}\left(x,y,z\right) & = & E\left(x,y,z\right)e^{-ikz},
\label{eq:electric field}
\end{eqnarray}
with $k$ being the wave number.  
Substituting Eq.$\,$(\ref{eq:electric field})
into Eq.$\,$(\ref{eq:Helmholtz}), we have
\begin{eqnarray}
\frac{\partial^{2}E}{\partial x^{2}}+\frac{\partial^{2}E}{\partial y^{2}}+\frac{\partial^{2}E}{\partial z^{2}}-2ik\frac{\partial E}{\partial z} & = & 0.
\end{eqnarray}
If the spreading rate in the propagation direction is much slower than that in the transverse plane, we can introduce the paraxial approximation condition
\begin{eqnarray}
\left|\frac{\partial^{2}E}{\partial z^{2}}\right| & \ll & \left|\frac{\partial^{2}E}{\partial x^{2}}\right|,\left|\frac{\partial^{2}E}{\partial y^{2}}\right|,\left|2k\frac{\partial E}{\partial z}\right|,
\end{eqnarray}
and rewrite the wave equation as
\begin{eqnarray}
2ik\frac{\partial E}{\partial z} & = & \frac{\partial^{2}E}{\partial x^{2}}+\frac{\partial^{2}E}{\partial y^{2}}.
\label{eq:paradiffraction}
\end{eqnarray}

\subsection{Fock-space optics}

We consider a bosonic mode subject to a coherent drive described by  $H=\varepsilon_\mathrm{p}\left(a+a^{\dagger}\right)$.
The quantum state of the mode is written as $\ket{\psi}=\sum_{n}c_{n}|n\rangle$.
Dynamics of the quantum state is governed by the Schr\"{o}dinger equation:
\begin{eqnarray}
i\frac{d}{dt}\ket{\psi} & = & H\ket{\psi}.
\end{eqnarray}
For the coefficient $c_{n}$ of the photon-number state, we have
\begin{eqnarray}
i\frac{d}{dt}c_{n}-\varepsilon_\mathrm{p}\sqrt{n+1}c_{n+1}-\varepsilon_\mathrm{p}\sqrt{n}c_{n-1} & = & 0.
\end{eqnarray}
When considering a quantum state with photon numbers $n\gg1$, the
nonuniform coupling rate is flattened, i.e., $\sqrt{n+1}\approx\sqrt{n}$.
Then, we have
\begin{eqnarray}
i\frac{d}{dt}c_{n}(t)-\varepsilon_\mathrm{p}\sqrt{n}c_{n+1}(t)-\varepsilon_\mathrm{p}\sqrt{n}c_{n-1}(t) & = & 0.\label{eq:dyn-nspace-2}
\end{eqnarray}
This differential equation can be conveniently solved by introducing
the Fourier transform between $n$ and $k$:
\begin{eqnarray}
C_{k}(t) & = & \frac{1}{\sqrt{2\pi}}\sum_{n}c_{n}(t)e^{-ink},\\
c_{n}(t) & = & \frac{1}{\sqrt{2\pi}}\int_{-\pi}^{\pi}C_{k}(t)e^{ink}dk.
\end{eqnarray}
Substituting the second equation into the dynamical equation Eq.$\,$($\text{\ref{eq:dyn-nspace-2}}$)
in the $n$-space, we obtain the system dynamics in the $k$-space:
\begin{eqnarray}
\left(i\frac{d}{dt}-2\sqrt{n}\varepsilon_\mathrm{p}\cos(k)\right)C_{k}(t) & = & 0.\label{eq:dyn-pspace}
\end{eqnarray}
The solution is
\begin{eqnarray}
C_{k}(t) & = & C_{k}(0)e^{-2i\sqrt{n}\varepsilon_\mathrm{p}\cos(k)t}.\label{eq:pspace-solution}
\end{eqnarray}
Transforming to the $n$-space, we have
\begin{eqnarray}
c_{n}(t) & = & \frac{1}{\sqrt{2\pi}}\int_{-\pi}^{\pi}C_{k}(t)e^{ink}dk\nonumber \\
 & = & \frac{1}{\sqrt{2\pi}}\int_{-\pi}^{\pi}C_{k}(0)e^{-2i\sqrt{n}\varepsilon_\mathrm{p}\cos(k)t}e^{ink}dk.
\end{eqnarray}

Equation (\ref{eq:dyn-nspace-2}) can also be written as
\begin{eqnarray}
i\frac{d}{dt}c_{n}(t) & = & \sqrt{n}\varepsilon_\mathrm{p}\Delta_{2}c_{n}(t)+2\sqrt{n}\varepsilon_\mathrm{p}c_{n}(t).\label{eq:adyn-nspance}
\end{eqnarray}
Here,
\begin{equation}
    \Delta_{2}c_n = c_{n+1}+c_{n-1}-2c_n,
\end{equation}
which is the second-order difference of the discrete variable $n$, corresponding to the second-order diffusion
terms $\frac{\partial^2}{\partial x^2}$ and $\frac{\partial^2}{\partial y^2}$ in Eq.~(\ref{eq:paradiffraction}). 

To see more explicitly the corresponding relationship between Eq.\,(\ref{eq:adyn-nspance}) for discrete variable $n$ and the second-order diffusion equation of Eq.\,(\ref{eq:paradiffraction}) for continuous variables, we examine quantum states with broad $n$-distributions and $c_n$ of smooth envelopes. In this case, the discrete variable $n$ can be treated as quasi-continuous. For example, coherent states with a broad $n$-distribution exhibit a narrow $k$-distribution, and thus we can make the approximation $\cos(k)\approx1-k^{2}/2$. Substituting this approximation into Eq.$\,$($\text{\ref{eq:dyn-pspace}}$), we have 
\begin{eqnarray}
\left[i\frac{d}{dt}-2\sqrt{n}\varepsilon_\mathrm{p}\left(1-\frac{k^{2}}{2}\right)\right]C_{k}(t) & = & 0.
\label{eq:approx}
\end{eqnarray}
Before transforming back to the $n$-space, we make the approximation $k\approx \sin(k)$ under the condition $k\ll 1$. Then, we have
\begin{equation}
\frac{k^2}{2}\approx 1-\cos k=-\frac{e^{ik}+e^{-ik}-2}{2}.
\label{eq:kSquareApprox}
\end{equation}
Substituting Eq.~(\ref{eq:kSquareApprox}) to Eq.~(\ref{eq:approx}) and transforming back to the $n$-space, we finally arrive at
\begin{eqnarray}
i\frac{d}{dt}c_{n}(t) & \approx & \sqrt{\bar{n}}\varepsilon_\mathrm{p}\frac{\partial^{2}}{\partial n^{2}}c_{n}(t)+2\sqrt{\bar{n}}\varepsilon_\mathrm{p}c_{n}(t).\label{eq:adyn-nspace}
\end{eqnarray}
Under the approximation $\sqrt{n}\approx\sqrt{\bar{n}}$, Eq.~(\ref{eq:adyn-nspace}) takes the same form as the paraxial
diffraction equation of optical beams, differing only by a global phase factor on the quantum state.

\subsection{Prism and lens in Fock space}
Starting from an initial coherent state, we discuss how prisms and lenses are constructed in the Fock space. In the following derivations, we neglect the global phase term in Eq.~(\ref{eq:adyn-nspace}). For large photon numbers, the Poisson distribution of a coherent state can be approximated by a Gaussian function:
\begin{equation}
    c_{\text{in}} = \frac{1}{(2\pi\sigma^2)^{1/4}} \exp\left[-\frac{(n-n_0)^2}{4\sigma^2}\right],
\end{equation}
where $\sigma=\sqrt{n_0}$ corresponds to the width of a Gaussian beam. 

\emph{Prism.} When the initial state is modified by a linear phase factor $k_0(n-n_0)$, the state becomes 
\begin{equation}
    c_{n}(0) = \frac{1}{(2\pi\sigma^2)^{1/4}} \exp\left[-\frac{(n-n_0)^2}{4\sigma^2}\right]\exp\left[ik_0(n-n_0)\right].
    \label{eq:linearphase}
\end{equation}
In momentum space, this linear phase shifts the distribution of the wavefunction,  centering it at $k=\Delta$ instead of $k=k_0$, which corresponds to an initial ``momentum kick". This imparts a constant velocity to the entire wavefunction, causing it to drift. By substituting the distribution [Eq.~(\ref{eq:linearphase})] to the wave equation [Eq.~(\ref{eq:adyn-nspace})], we find the solution to be a moving and spreading Gaussian:
\begin{equation}
    P(n,t)=|c_n(t)|^2=\frac{1}{\sqrt{2\pi\sigma(t)^2}}\exp\left[-\frac{(n - n_c(t))^2}{2\sigma(t)^2}\right],
    \label{eq:refraction}
\end{equation}
where
\begin{equation}
    n_c(t) = n_0 + v_gt=n_0-2k_0\sqrt{n_0}\varepsilon_\mathrm{p} t
\end{equation}
is the moving position of the wavefunction center and 
\begin{equation}
    \sigma(t)^2=\sigma^2+\left( \frac{\sqrt{n_0}\varepsilon_\mathrm{p} t}{\sigma} \right)^2
\end{equation}
is the spreading width of the wavefunction.
In the short time limit $\bar{n}\varepsilon_\mathrm{p} t\ll \sigma$, the spreading of the photon-number distribution can be neglected, leaving a notable refraction effect with the refraction angle $\theta$ determined by $\tan \theta=-2k_0\sqrt{n_0}\varepsilon_\mathrm{p}$.

\emph{Lens.} When the initial state is modified by a quadratic phase factor centered at $n_0$, the state becomes
\begin{equation}
        c_{n}(0) = \frac{1}{(2\pi\sigma^2)^{1/4}} \exp\left[-\frac{(n-n_0)^2}{4\sigma^2}\right]\exp\left[i \phi_0 (n-n_0)^2\right].
\end{equation}
Substituting this wavefunction into Eq.~(\ref{eq:adyn-nspace}), the solution remains a Gaussian function with a stationary center and a dynamic width:
\begin{equation}
    P(n,t)=|c_n(t)|^2=\frac{1}{\sqrt{2\pi\sigma(t)^2}}\exp\left[-\frac{(n - n_0)^2}{2\sigma(t)^2}\right],
    \label{eq:lens}
\end{equation}
where
\begin{equation}
    \sigma(t)^2=\sigma^2\left[ (1 - 4\sqrt{n_0}\varepsilon_\mathrm{p}\phi_0 t)^2 + \left(\frac{\sqrt{n_0}\varepsilon_\mathrm{p}t}{\sigma^2}\right)^2 \right]
\end{equation}
is the time-dependent width of the photon-number distribution. The evolution of the width becomes a parabolic function of time, governed by the interplay between the parabolic phase coefficient $\phi_0$ and initial width $\sigma=\sqrt{n_0}$. It can be easily deduced that the Gaussian wavefunction reaches a minimum width
\begin{equation}
  \sigma_\mathrm{min}^2 = \frac{\sigma^2}{1+16\phi_0^2\sigma^4} \approx \frac{1}{16\phi_0^2 \sigma^2},
  \label{eq:focalpoint}
\end{equation}
at a special focal time
\begin{equation}
    t_\mathrm{min} = \frac{\phi_0}{\sqrt{n_0}\varepsilon_\mathrm{p}(4\phi_0^2+1/4\sigma^4)} \approx \frac{1}{4\sqrt{n_0}\varepsilon_\mathrm{p}\phi_0}.
\end{equation}
For the case $\phi_0>0$ and $t_\mathrm{min}>0$, the width of the distribution first narrows to its minimum for $t<t_\mathrm{min}$ and then spreads after the focal time, corresponding to a convex lens in optics. In contrast, for $\phi_0<0$ and $t_\mathrm{min}<0$, the width of the distribution continuously increases, corresponding to a concave lens. 
According to Eq.~(\ref{eq:focalpoint}), the minimum width is inversely proportional to the width of the initial state, which means that the initial and focused states are linked by a Fourier transform up to a scaling factor $4\phi_0$. In practice, the paraxial condition also limits the minimum width of the focused state, which requires $\phi_0\sigma<1$ for the validity of the wave equation [Eq.~(\ref{eq:adyn-nspace})] and results in a bounded width of the focused state $\sigma_\mathrm{min}^2>1$.

\section{Phase-space slingshot method}

\subsection{Principle of the phase-space slingshot method}

\begin{figure}[t]
    \centering
    \includegraphics{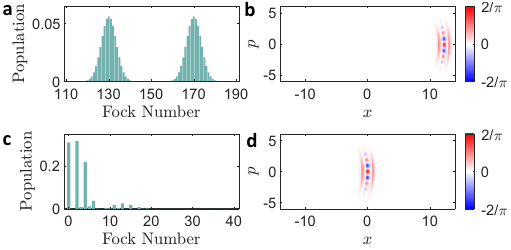}
    \caption{\textbf{Phase-space slingshot method to prepare a double-Gaussian (DG) state.}  
    \textbf{a, b, } Fock-space population distribution and phase-space Wigner function of a target DG state $\ket{\psi_\mathrm{DG}(n_1=130,r_1=1,n_2=170,r_2=1,\theta=0)}$.
    \textbf{c, d, } Fock-space population distribution and phase-space Wigner function of the displaced target state $\hat{D}^\dagger(\beta) \ket{\psi_\mathrm{DG}}$, with the displacement amplitude $\beta =\sqrt{150}$.
    Direct preparation of $\ket{\psi_\mathrm{DG}}$ with a GRAPE pulse is challenging due to the state's high dimension. However, $\hat{D}^\dagger(\beta) \ket{\psi_\mathrm{DG}}$ is much easier and practicable to prepare, because it is localized near the phase-space origin and has a much lower photon distribution. Therefore, in experiment we first use GRAPE to prepare $\hat{D}^\dagger(\beta) \ket{\psi_\mathrm{DG}}$ and then apply a displacement operation $\hat{D}(\beta)$ to obtain the target state.  
    }
    \label{fig:pssm}
\end{figure}

Here we introduce a method called the phase-space slingshot to effectively prepare the double-Gaussian (DG) distributed state required for the experiments shown in Fig.~3 and Fig.~5 in the main text. 
We first provide a more specific definition of the DG state:
\begin{equation}
    \ket{\psi_\mathrm{DG}(n_1,r_1,n_2,r_2,\theta)} = \frac{1}{\sqrt{r_1^2+r_2^2}}[r_1\ket{G_{n_1}}+r_2e^{\mathrm{i}\theta}\ket{G_{n_2}}],
\end{equation}
where $n_1$, $n_2$ represent the centers of the two Gaussian peaks, $r_1:r_2$ is the ratio of the heights of the two peaks, $\theta$ is the phase between the two components, and $\ket{G_m}$ is defined as a Gaussian-distributed state
\begin{equation}
    \ket{G_m} = \frac{1}{A_\mathrm{G}}\sum_n \bigg\{ \exp \bigg [ {-\frac{(n-m)^2}{2\sigma^2}} \bigg ] \bigg \} \ket{n}, 
\end{equation}
where $n$ is the Fock number, $A_\mathrm{G}$ is the normalization parameter, and $\sigma = 5$ for all experiments.

Next, we introduce how to prepare such a state.
Taking $\theta=0,~r_1:r_2=1:1.~n_1=130,~n_2=170$ as an example, the target state's Fock-space distribution is shown in Fig.~\ref{fig:pssm}a. Given the exceptionally large photon numbers, it is not practical to prepare it directly using standard techniques such as GRAPE~\cite{UControlKhaneja2005JMRGRAPE,UControlHeeres2017NCGRAPE}. Instead, we examine its Wigner function in phase space, as shown in Fig.~\ref{fig:pssm}b. The Wigner function resembles a distorted squeezed cat state. Due to the distribution of the DG state over multiple Fock components, it does not occupy a large region in phase space as a single Fock state would. Instead, it remains localized near $(x,p)=(\sqrt{150},0)$, with amplitudes suppressed due to destructive interference away from this region. By displacing this state back to the origin (Fig.~\ref{fig:pssm}d), we effectively transform it into a low-photon-number state.
To verify this approach, we plot its Fock-space distribution in Fig.~\ref{fig:pssm}c. The state can be truncated to 35 dimensions, allowing us to prepare it efficiently using GRAPE within $2\,\mathrm{\mu s}$.

Experimentally, the DG state is synthesized in reverse. We first use GRAPE to prepare a truncated, distorted squeezed cat state near the origin in the phase space (Figs.~\ref{fig:pssm}c and \ref{fig:pssm}d). We then apply a fast displacement operation to shift it to $(x,p)=(\sqrt{150},0)$, which yields the DG state (Figs.~\ref{fig:pssm}a and \ref{fig:pssm}b). This process resembles launching a small-photon-number state into a large-photon-number region, akin to a slingshot, and is thus referred to as the phase-space slingshot.
Additionally, during this process, we can map the ground ($\ket{\mathrm{g}}$) and excited ($\ket{\mathrm{e}}$) states of the ancilla qubit to the left and right peaks of the DG state, respectively. This allows us to control the relative height and phase between the two peaks of the DG state by adjusting the amplitude ratio and relative phase of the $\ket{\mathrm{g}}$ and $\ket{\mathrm{e}}$ components in the qubit's initial state.

To further optimize this process, we incorporate the time-reversal dynamics of the displacement operation under self-Kerr and other high-order effects into the simulations. This refined model replaces the simple displacement operation in Fig.~\ref{fig:pssm}, and we employ its output as the GRAPE optimization target to enhance the overall preparation fidelity.

In the double-slit interference experiment, preparing narrower ``slits'' or increasing the slit separation presents significant challenges.
Increasing $d$ causes the distorted squeezed cat state to spread laterally, while decreasing $\sigma$ enhances squeezing, spreading the state broadly along one phase-space quadrature. Both effects significantly increase the truncation dimension required for accurate preparation, thereby complicating the GRAPE optimization.

\subsection{GRAPE target states involved in the phase-space slingshot method}

\begin{figure}[tbp]
    \centering
    \includegraphics{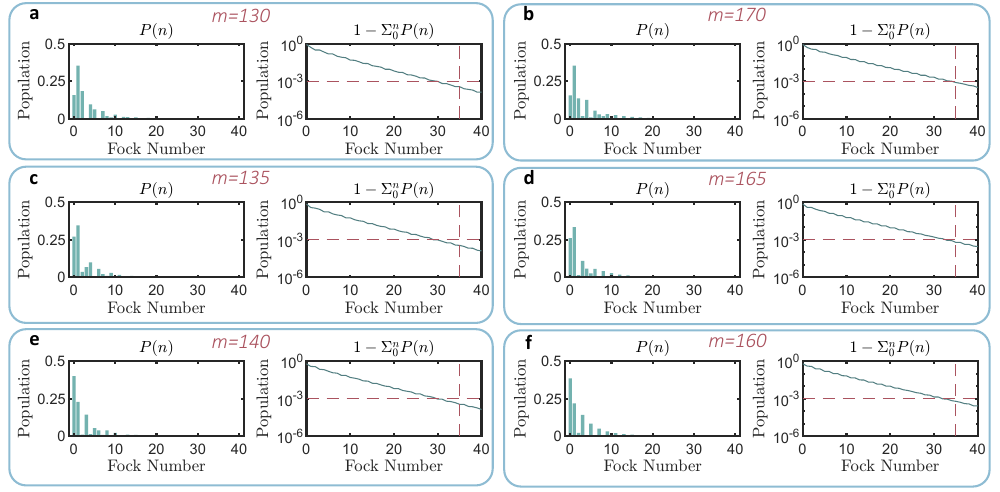}
    \caption{\textbf{Fock-space distributions of the target states for GRAPE pulse optimization.} \textbf{a-f}, Fock-space population distributions of the states $\hat{D}[-\sqrt{150}]\ket{G_m}$ for $m=130,~170,~135,~165,~140,~160$, respectively. These states serve as the target states for the GRAPE pulses in this work. In practice, the results are obtained through a time-reversed simulation (without dissipation) of the displacement process, taking into account the self-Kerr ($K_4$ and $K_6$) and cross-Kerr ($\chi$ and $K_\mathrm{e}$) coefficients defined in Eq.~(\ref{eq:Hamitonian}).  
    The right plots in each sub-figure display the cumulative populations outside the first $n$ dimensions. 
    The dash lines indicate that the cumulative population outside the first 35 dimensions is less than 0.1\% for all these states.
    }
    \label{fig:PSstate}
\end{figure}

Here we provide the GRAPE target states involved in experiments.
To prepare a DG-distributed state, we intend to realize an arbitrary superposition of two Gaussian-distributed states centered at $n_1$ and $n_2$, with relative magnitude ($r_1:r_2$) and phase ($\theta$).
To achieve this, we first prepare a superposition state $ \frac{1}{ \sqrt{ r_1^2 + r_2^2 } } \left[ r_1 \ket{g} + r_2 e^{\mathrm{i} \theta } \ket{e} \right] $ on the qubit. We then use a GRAPE-optimized control pulse and a fast displacement operation $ \hat{D}\left[\sqrt{\frac{n_1+n_2}{2}}\right]$ to map the quantum information on the qubit to the two Gaussian-distributed states in the cavity.
Consequently, the control of $r_1:r_2$ and $\theta$ is effectively translated into the initialization of the qubit. 
The GRAPE operation realizes the transformation as follows:
\begin{equation}
    \ket{g}\ket{0} \rightarrow \ket{g}\hat{D}^\dagger\left[\sqrt{\frac{n_1+n_2}{2}}\right]\ket{G_{n_1}}, \qquad  \ket{e}\ket{0} \rightarrow \ket{g}\hat{D}^\dagger\left[\sqrt{\frac{n_1+n_2}{2}}\right]\ket{G_{n_2}}.
\end{equation}

In our experiment, we use three sets of target states $\{n_1=130,\,n_2=170\}$, $\{n_1=135,\,n_2=165\}$, and $\{n_1=140,\,n_2=160\}$, as shown in Fig.~\ref{fig:PSstate}.
The cumulative populations outside the first 35 dimensions is less than 1\textperthousand~for all states in Figs.~\ref{fig:PSstate}a-f, with calculated values of 0.36\textperthousand, 0.77\textperthousand, 0.36\textperthousand, 0.62\textperthousand, 0.35\textperthousand, 0.61\textperthousand, respectively. 
This justifies truncating the GRAPE optimization targets to 35 dimensions.

\begin{figure}[tbp]
    \centering
    \includegraphics{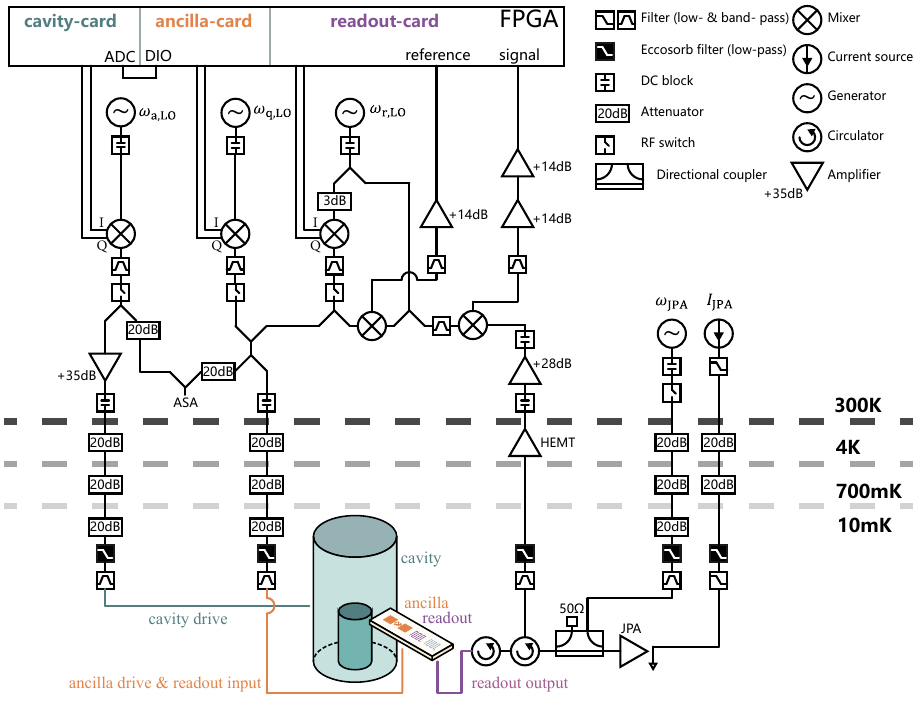}
    \caption{\textbf{Wiring diagram of the system.} A legend for the main elements is given in the upper right corner. Attenuation and gain values are indicated in dB. The temperatures for different plates of the refrigerator are indicated with dashed lines and labels on the right. 
    The control lines involved in the device are marked with different colors for clarity.
    }
    \label{fig:wiring}
\end{figure}

\section{Experimental setup}
\subsection{Wiring diagram and system parameters}
The wiring of the experiment is shown in Fig.~\ref{fig:wiring}.
All microwave control signals in the experiment are generated by the single-sideband technique with IQ mixers. The IQ signals are produced by Field Programmable Gate Array (FPGA) cards, which are triggered by the same clock card (not drawn). These signals are mixed with their respective local signals, filtered, routed through switches, and then sent into the fridge. In this work, three FPGA cards are used, labeled as {cavity-card}, {ancilla-card}, and {readout-card}, which provide sidebands of $-100 \,\mathrm{MHz}$, $-100 \,\mathrm{MHz}$, and $-50 \,\mathrm{MHz}$, respectively. Between the {cavity-card} and {ancilla-card}, an additional line is connected to transmit a digital signal from one card to the other. This setup tracks the delay between the two cards, ensuring the effectiveness of operations such as GRAPE pulses.

Experimental implementations sometimes require detuning the cavity or qubit drive frequency from resonance, such as the weak single-photon pump or frequency-selective qubit $\pi$-pulses. 
In both the main text and the Supplementary Materials, resonant and detuned pulses in the sequence diagrams are color-coded as follows: green for resonant cavity drives, blue for detuned cavity drives, orange for resonant qubit drives, and yellow for detuned qubit drives.

\begin{table}
\begin{center}

\begin{tabular}{l|l|c|c|l} \hline \hline
\textbf{Mode} & \textbf{Parameter} & \textbf{Symbol} & \textbf{Value} & \textbf{Method}\\ \hline \hline

\multirow{8}{*}{Resonator (3D cavity)} 
& frequency & $\omega_\mathrm{a}/2\pi$ & $6.140\,\mathrm{GHz}$ & Ramsey experiment for cavity\\ \cline{2-5}
& 4th-order self-Kerr & $K_\mathrm{4}/2\pi$ & $2.18\,\mathrm{kHz}$ & Ramsey at small photon numbers\\ \cline{2-5}
& 6th-order self-Kerr & $K_\mathrm{6}/2\pi$ & $1\,\mathrm{Hz}$ & Fock-space spectrometer\\ \cline{2-5}
& single-photon decay time & $T_\mathrm{1,c}$ & $1.6\,\mathrm{m s}$ & Standard coherence measurement \\ \cline{2-5}
&  transverse relaxation time & $T_\mathrm{2,c}$ & $2.0\,\mathrm{m s}$ & Standard coherence measurement \\ \cline{2-5}
& thermal & $n_\mathrm{th,c}$ & 3\% & Purification \\ \cline{2-5}
& Kerr-e when the ancilla is excited & $K_\mathrm{e}/2\pi$ & $0.52\,\mathrm{kHz}$ & Wide-range number splitting experiment \\ \cline{2-5}
& cross-Kerr with ancilla & $\chi/2\pi$ & $0.596\,\mathrm{MHz}$ & Wide-range number splitting experiment \\ \hline

\multirow{9}{*}{Ancilla (transmon qubit)} 
& frequency & $\omega_\mathrm{q}/2\pi$ & $5.290\,\mathrm{GHz}$ & Rabi and Ramsey experiment\\ \cline{2-5}
& Kerr & $K_\mathrm{q4}/2\pi$ & $153\,\mathrm{MHz}$ & Sweep $\ket{\mathrm{g}}\leftrightarrow\ket{\mathrm{f}}$ transition\\ \cline{2-5}
& single-photon decay time & $T_\mathrm{1,q}$ & $80\,\mathrm{\mu s}$ & Standard coherence measurement \\ \cline{2-5}
& transverse relaxation time & $T_\mathrm{2,q}$ & $110\,\mathrm{\mu s}$ & Standard coherence measurement \\ \cline{2-5}
& transverse relaxation time (echo) & $T_\mathrm{2E,q}$ & $130\,\mathrm{\mu s}$ & Standard coherence measurement \\ \cline{2-5}
& thermal & $n_\mathrm{th,q}$ & 0.7\% & Purification \\  \cline{2-5}
& readout frequency & $\omega_\mathrm{r}/2\pi$ & $7.590\,\mathrm{GHz}$ & Direct RF measurement\\ \cline{2-5}
& readout resonator linewidth  & $\kappa_\mathrm{r}/2\pi$ & $2.8\,\mathrm{MHz}$ & Fitting the readout curve\\ \cline{2-5}
& cross-Kerr with its readout resonator & $\chi_\mathrm{qr}/2\pi$ & $2.5\,\mathrm{MHz}$ & Fitting the readout curve\\ \hline

\end{tabular}
\caption{\textbf{Main Parameters.} A summary of the main system parameters used in this work. 
}
	\label{tab:allparameters}

\end{center}
\end{table}

The key parameters of the whole system are displayed in Table~\ref{tab:allparameters}, including the frequencies and coherence properties of the cavity and qubit, the self-Kerr coefficients of the cavity, and coupling strengths between neighborhood elements.

\begin{figure}[tbp]
    \centering
    \includegraphics{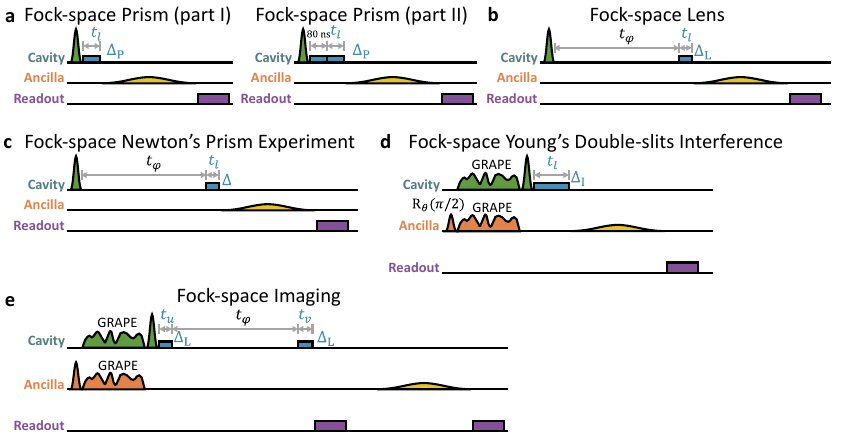}
    \caption{\textbf{Experiment sequences used in the main text.} 
    Green, orange, and purple segments represent resonant drives for the cavity, ancilla, and readout modes, respectively. Blue segments represent the weak single-photon pump on the cavity with a small detuning if needed.
    Yellow segments represent the frequency-selective $\pi$ pulses applied to the ancilla. The frequencies of these pulses, which correspond to different Fock states in the cavity, are determined by the Fock-space camera (see Fig.~1e in the main text). Purification processes before the experiment are not drawn for all sub-figures for clarity.
    \textbf{a,} Sequence for Figs.~2a-d in the main text.
    Here, the detuning of the weak single-photon drive is set to $\Delta_\mathrm{P}/2\pi=0.5\,\mathrm{MHz}$.
    Part I and part II correspond to the left and right parts of Figs.~2c and 2d, respectively.
    In part II, we change the phase of the second single-photon pump segment by $\Delta_\mathrm{P}\delta t=\pi/2$, which is equivalent to letting the state evolve under the given detuning for $\delta t=500\,\mathrm{ns}$ and accumulate a linear phase. 
    \textbf{b,} Sequence for Fig.~2e-k in the main text.
    After preparing a coherent state, we let the state undergo free evolution under the self-Kerr for $t_\varphi=4684\,\mathrm{ns}$, followed by a weak single-photon pump. During the waiting period, the combined effects of self-Kerr and detuning $\Delta_\mathrm{L}/2\pi=0.33\,\mathrm{MHz}$ induce a quadratic phase accumulation on the quantum states, which centers around $n=150$ and is similar to a classical lens.
    \textbf{c,} Sequence for Fig.~3 in the main text.
    The difference between \textbf{c} and \textbf{b} is the detuning of the weak single-photon pump.
    \textbf{d,} Sequence for Fig.~4 in the main text. 
    The first three parts (a single qubit gate, GRAPE pulses, and a strong displacement on the cavity) form the phase-space slingshot. 
    The interference pattern exhibits some curvature due to the self-Kerr effect. 
    We compensate the curvature with a detuning $\Delta_\mathrm{I}/2\pi=0.29\,\mathrm{MHz}$, so that the Fock-space distribution remains near $n = 150$ throughout the process.
    \textbf{e,} Sequence for Fig.~5 in the main text.
    The state preparation uses the same phase-space slingshot method.  
    The parameters $t_\varphi$ and $\Delta_\mathrm{L}$ are calibrated in \textbf{b}.
    }
    \label{fig:allseq}
\end{figure}

\subsection{Experiment sequences used in the main text}

To focus on the underlying physics rather than on detailed implementations, we present simplified schematic sequences in the main text when introducing various Fock-space optical devices and applications. 
These sequences are designed to highlight general principles. 
For completeness, here we provide more operational pulse sequences that directly show the envelopes of microwave drives applied to each mode, as shown in Fig.~\ref{fig:allseq}.  
Green, orange, and purple indicate resonant drives; blue and yellow indicate off-resonant drives. 
The blue pulses represent the weak single-photon drive applied to the cavity, which serves as the core evolution mechanism of ``light" in Fock space in our experiment. A detuning is generally applied on the blue pulses to control the linear phase accumulation.
The yellow pulses typically correspond to selective $\pi$ pulses on the ancilla qubit, with a detuning according to the target Fock states.

In Fig.~\ref{fig:allseq}, the gate durations are set as follows. First, Gaussian-shaped gates typically have durations of four standard deviations. The cavity displacement gate has a duration of $160\,\mathrm{ns}$, which, under current self-Kerr nonlinearity and power constraints, can displace the cavity to approximately 1000 photons, more than sufficient for the 100-level operations performed in this experiment. For qubit control, we employ two types of gates: a fast, broadband excitation gate ($200\,\mathrm{ns}$) for general operations and reset (Fig.~\ref{fig:rabi}), and a slow, frequency-selective gate ($3\,\mathrm{\mu s}$) for probing the population of specific Fock states in the cavity~\cite{NumSplSchuster2007Nature}. 
The readout duration is fixed at $1\,\mathrm{\mu s}$, followed by a $1\,\mathrm{\mu s}$ wait for photon leakage from the readout resonator. Details of the readout performance can be found in Sec.~\ref{sec:histogram}. Finally, the parity measurement in the initial purification process (not drawn) includes a wait time of $596\,\mathrm{ns}$.

\section{Calibrations at large photon numbers} 
\subsection{Readout at large photon numbers}\label{sec:histogram}

\begin{figure}[htbp]
    \centering
    \includegraphics{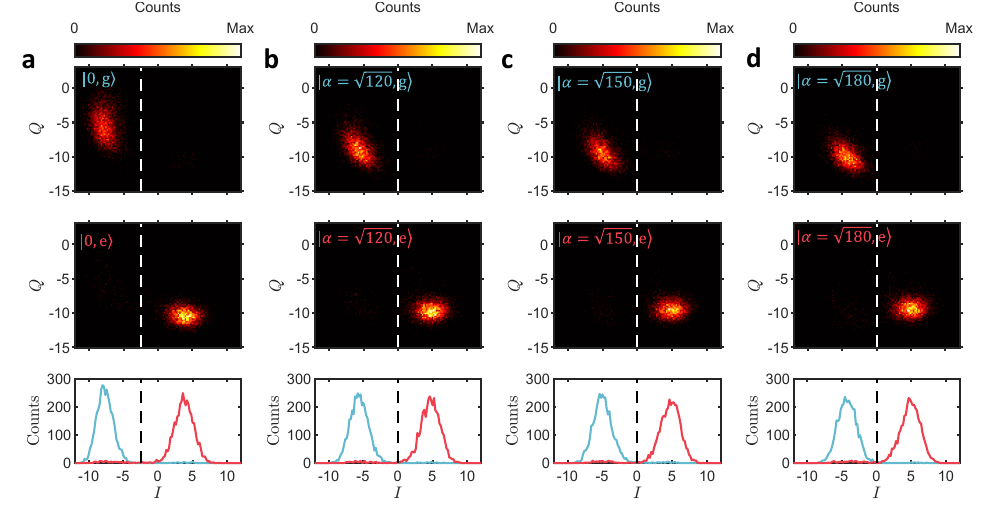}
    \caption{\textbf{Readout histograms at zero and large photon numbers.} 
    We first initialize the qubit to either $\ket{\mathrm{g}}$ or $\ket{\mathrm{e}}$, and then either resonantly drive the cavity to a coherent state with different average photon numbers or leave it in the vacuum state. We get the readout histograms with the average cavity photon number being  0 (\textbf{a}), 120 (\textbf{b}), 150 (\textbf{c}), 180 (\textbf{d}), respectively.
    In \textbf{a-d}, the top and middle panels show the histogram results after preparing qubit to $\ket{\mathrm{g}}$ or $\ket{\mathrm{e}}$, respectively, while the bottom panel shows the projected count distribution along the $I$ axis.
    The $I$ and $Q$ axes represent the demodulated readout signal from the FPGA and are proportional to voltage. 
    The dashed lines indicate the threshold on the $I$ axis used to distinguish the qubit states $\ket{\mathrm{g}}$ and $\ket{\mathrm{e}}$. 
    The thresholds in \textbf{b-d} are identical, demonstrating that a fixed threshold is valid for photon numbers in the range of 120 to 180.
    }
    \label{fig:hist}
\end{figure}

Our system is a cavity-qubit dispersive coupling system~\cite{CQEDWallraff2004Nature,CQEDBlais2021RMP}, where the qubit is also dispersively coupled to a readout resonator. 
We perform a dispersive readout to determine the state of the qubit, as shown in Fig.~\ref{fig:hist}a.
Typically, the cross-Kerr coupling between the cavity and the readout resonator is negligible compared to the qubit-readout dispersive shift. However, with large cavity photon numbers, this cross-Kerr coupling becomes more pronounced, introducing a frequency shift on the readout resonator. This cavity-induced readout frequency shift has been rarely discussed in existing studies.

To calibrate the readout, we first scan the transmission spectrum of the readout resonator with the cavity in a large coherent state ($\ket{\alpha=\sqrt{150}}$). 
Then we set the readout probe frequency to the midpoint between the peaks corresponding to the joint states $\ket{\alpha=\sqrt{150},\mathrm{g}}$ and $\ket{\alpha=\sqrt{150},\mathrm{e}}$.
The readout histograms at this probe frequency are shown in Figs.~\ref{fig:hist}b-d. The required initial quantum states in these experiments are prepared by first initializing the qubit and then displacing the cavity. Note that when the qubit is in $\ket{\mathrm{e}}$, the displacement frequency should be adjusted by $\chi$. Compared to Fig.~\ref{fig:hist}a, the overall rotation in Figs.~\ref{fig:hist}b-d is attributed to the cross-Kerr coupling between the cavity and the readout resonator.

Strictly speaking, measuring different Fock states in our experiments requires different readout thresholds. However, within our operational range with large photon numbers (120-180), the histograms are sufficiently separated, besides the rotation of the readout signal in the $I-Q$ plane caused by the cross-Kerr coupling is small enough. As a result, a single threshold suffices, as demonstrated in Figs.~\ref{fig:hist}b-d. 

Therefore, we employ only two thresholds in this work, except for the large-range number-splitting experiment presented in Fig.~1e in the main text. We use the threshold from Fig.~\ref{fig:hist}a for low-photon-number cavity states (e.g., during initial purification), and that from Figs.~\ref{fig:hist}b-d for high-photon-number cavity states (e.g., before and after the qubit frequency-selective $\pi$ pulses).

\subsection{Fock-space camera}\label{sec:camera}
To obtain a ``camera" that maps Fock states to their corresponding qubit frequencies, we prepare coherent states of varying sizes using different drive strengths and perform number-splitting experiments near the center of each coherent state. 
The displacement strength for each experiment is chosen such that the highest-photon-number signal from one experiment reappears in the next, enabling reliable Fock-state counting from zero upward.
Example spectra are shown in the insets of Fig.~1e in the main text.
The minimum frequency step for the scans is $0.05\,\mathrm{MHz}$, providing sufficient precision for our experimental system. Based on Eq.~(\ref{eq:Hamitonian}), the data are fitted to the relation between the qubit frequency and the corresponding Fock number in the cavity, i.e., $\Delta_\mathrm{q}(n) = -n\chi-n(n-1)\frac{K_\mathrm{e}}{2}$, where $\Delta_\mathrm{q}(n)=\omega_\mathrm{q}(n)-\omega_\mathrm{q}(0)$ is the qubit frequency shift when the cavity is in Fock state $\ket{n}$.
As shown in Fig.~1e in the main text, a quadratic fit suffices, confirming that higher-order terms are negligible within our photon-number range ($n\approx180$).

\subsection{Rabi and Ramsey experiments with the cavity in large Fock states}

\begin{figure}[htbp]
    \centering
    \includegraphics{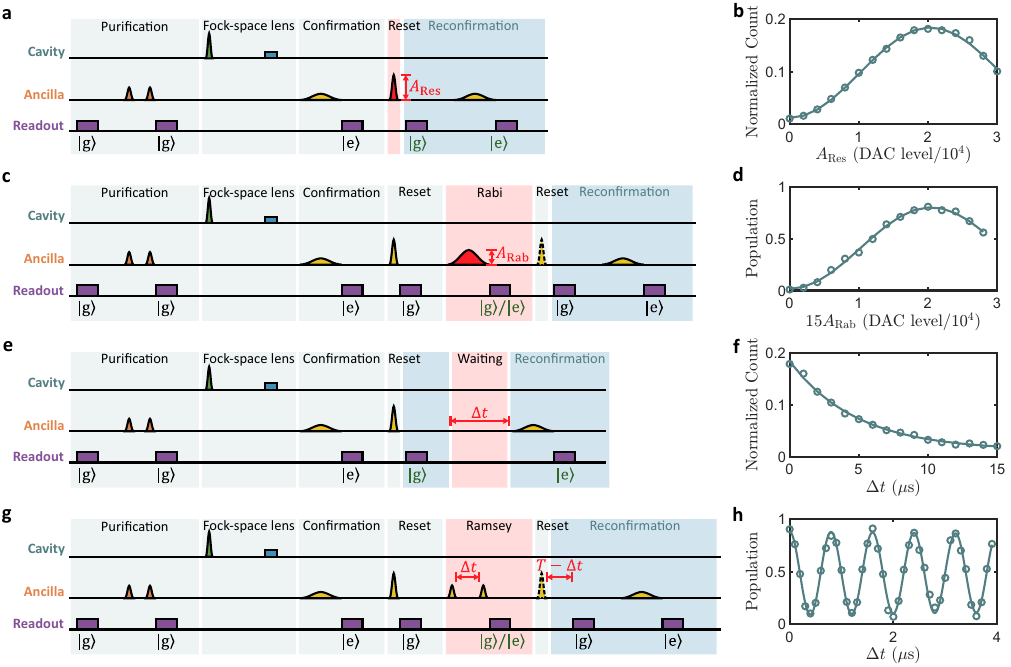}
    \caption{\textbf{Rabi experiment and cavity $T_{1,\mathrm{c}}$ measurements with the cavity at large Fock states.} 
    \textbf{a, b,} Sequence and experiment results of the Rabi experiment used to calibrate the $\pi$ pulse of the qubit when the cavity is at a large Fock state.
    The system is first purified, followed by the application of a Fock-space lens and a confirmation measurement to prepare the cavity in the Fock state $\ket{n=150}$. A Gaussian-shaped pulse with varying amplitudes is then applied to reset the qubit to $\ket{\mathrm{g}}$, and a final reconfirmation step ensures the cavity remains in $\ket{n=150}$.
    Only in \textbf{a}, the frequency-selective $\pi$ pulse used for confirmation and
    reconfirmation is temporarily set with standard qubit $\pi$ pulse strength calibrated at Fock \ket{0}.
    In \textbf{b}, the results are defined as the normalized count $N(g,g,e,g,e)/N(g,g,e,a,a)$, where $N$ is the count of the data and its argument corresponds to the five readout results in \textbf{a}, with $g$, $e$, and $a$ denoting the ground state, excited state, and all outcomes, respectively.
    \textbf{c, d,} Sequence and experiment results of the Rabi experiment used to calibrate the combined fidelity of the frequency-selective $\pi$ pulse and the readout with the cavity at a large Fock state. The amplitude of the reset pulse has already been calibrated in \textbf{a}. A second reset is applied if the qubit is measured to be in the $\ket{\mathrm{e}}$ state in the Rabi experiment. In \textbf{d}, the results are defined as: $N_2(g,g,e,g,e,g,e)/[N_1(g,g,e,g,g,g,e)+N_2(g,g,e,g,e,g,e)]$, where $N_1$ and $N_2$ denote the counts for the cases without and with the second reset, respectively.
    \textbf{e, f,} Sequence and experiment results of measuring $T_{1,\mathrm{c}}$ when the cavity is at a large Fock state.
    In \textbf{f}, the results are defined as: $N(g,g,e,g,e)/N(g,g,e,a,a)$.
   \textbf{g, h,} Sequence and experiment results of the Ramsey experiment used to calibrate the qubit frequency and evaluate its coherence. The second reset is conditional on the Ramsey experiment outcome. In \textbf{h}, the results are defined as $N_2(g, g, e, g, e, g, e)/[N_1(g, g, e, g, g, g, e) + N_2(g, g, e, g, e, g, e)]$. In \textbf{b}, \textbf{d}, \textbf{f}, and \textbf{h}, circles represent the experimental data and lines are fits (cosine, exponential, or exponentially decaying cosine).
    }
    \label{fig:rabi}
\end{figure}

The presence of a large photon number (e.g., Fock state $\ket{n=150}$) in the cavity shifts the qubit frequency by tens of MHz, necessitating recalibration of the qubit $\pi$ pulse under such conditions for accurate photon-number distribution measurements. The pulse sequence for a qubit Rabi experiment with the cavity in a large Fock state is shown in Fig.~\ref{fig:rabi}a. The Fock-space lens technique is used to efficiently prepare the $\ket{n=150}$ Fock state. To ensure the cavity remains in this high-photon-number state, we perform confirmation and reconfirmation checks before and after the experiment, respectively. A qubit flip following the frequency-selective $\pi$ pulse in both checks confirms that the cavity has not decayed. For these checks, the frequency-selective $\pi$ pulse strength is initially set to a standard qubit $\pi$ pulse calibrated at Fock $\ket{0}$. Figure~\ref{fig:rabi}a is to calibrate a narrow $\pi$ pulse (used for reset) first, which can then be scaled accordingly to generate wide, frequency-selective $\pi$ pulses. The results of the Rabi oscillation measurement are shown in Fig.~\ref{fig:rabi}b, presented as normalized counts $N(g,g,e,g,e)/N(g,g,e,a,a)$. Here, the argument of $N$ lists the readout outcomes ($g$: ground, $e$: excited, $a$: all, regardless of $g$ or $e$) throughout the whole experimental sequence. A fit to the data reveals that the optimal $\pi$ pulse strength for $\ket{n=150}$ differs by 6\% compared to that for Fock $\ket{0}$.

Using the $\pi$ pulse calibrated in Figs.~\ref{fig:rabi}a and \ref{fig:rabi}b, we further estimate the fidelity in our method for measuring the Fock state population. The pulse sequence for this experiment is shown in Fig.~\ref{fig:rabi}c. Between the confirmation and reconfirmation stages, we vary the strength of the frequency-selective $\pi$ pulse to rotate the qubit, while the cavity remains in $\ket{n=150}$. If the qubit is subsequently measured in $\ket{\mathrm{e}}$ after this Rabi experiment, a second reset is applied before reconfirmation. The results shown in Fig.~\ref{fig:rabi}d are given by $N_2(g, g, e, g, e, g, e)/[N_1(g, g, e, g, g, g, e) + N_2(g, g, e, g, e, g, e)]$, where $N_1$ and $N_2$ denote the counts for the cases without and with the second reset, respectively. The fit indicates an upper bound of 80\% and a lower bound of 1.6\% for the cavity readout. 

Another application of our confirmation-reconfirmation protocol is to measure the decay rate of large Fock states. Theoretically, the lifetime of a Fock state $\ket{n}$ scales as $T_{1,\mathrm{c}}(n) = T_{1,\mathrm{c}}(1)/n$~\cite{FockWang2008PRL}, which predicts a lifetime of $1.6\,\mathrm{ms}/150 = 11\,\mathrm{\mu s}$ for $\ket{n=150}$ state in our system. The measurement sequence is shown in Fig.~\ref{fig:rabi}e. The normalized counts $N(g,g,e,g,e)/N(g,g,e,a,a)$ are shown in Fig.~\ref{fig:rabi}f, which represent the remaining population of $\ket{n=150}$. A fit to the data yields $T_{1\mathrm{c}}=4.8\,\mathrm{\mu s}$, significantly shorter than the theoretical value. This discrepancy might arise because the transmon coupled to the cavity can carry a portion of the energy when in a dressed state. Furthermore, an unused qubit coupled to the same cavity might have unwanted transitions, potentially leading to transmon ionization~\cite{HPhotonLucas2019PRApplied,HPhotonLescanne2019PRApplied,ChaosCohen2023PRXQTransmon,ChaosDumas2024PRXIonization}, which would unexpectedly reduce the cavity's effective lifetime.

The Ramsey experiment is used to calibrate the qubit frequency and evaluate its coherence. 
Our confirmation-reconfirmation protocol can also be applied to perform a qubit Ramsey experiment. As shown in Fig.~\ref{fig:rabi}e, a standard qubit Ramsey experiment is conducted between the confirmation and reconfirmation stages, followed by a reset conditional on the Ramsey measurement outcome. The results shown in Fig.~\ref{fig:rabi}h are defined as $N_2(g, g, e, g, e, g, e)/[N_1(g, g, e, g, g, g, e) + N_2(g, g, e, g, e, g, e)]$. The extracted qubit frequency agrees well with the value obtained from the number-splitting experiment, differing by only $-0.2\pm 2.9$~kHz. The results also confirm that qubit coherence is well maintained over the duration of the Ramsey evolution.

The results in Figs.~\ref{fig:rabi}d and \ref{fig:rabi}f can partially explain the discrepancies observed in Figs.~2g and 2h of the main text. First, after correcting for the readout infidelity due to qubit decoherence (Fig.~\ref{fig:rabi}d), the actual population at the focal point should be around $17\%/80\%=21\%$. Second, the shortened cavity lifetime (Fig.~\ref{fig:rabi}f) would further reduce the Fock state population during focusing.
A full quantitative model is challenging because the cavity's coherence time $T_{2,\mathrm{c}}$ at large photon numbers is difficult to measure.
However, if we assume that $T_{2,\mathrm{c}}$ is comparable to $T_{1,\mathrm{c}}$, the simulation that excludes readout errors matches our experimental data well. Therefore, we conclude that the primary sources of the discrepancy between the experimental results of the Fock-space lens and the simulation results in the main text are the measurement errors and the reduced cavity lifetime.

\section{Supplementary data and curve fitting}

In this section, we present supplementary experimental data and relevant fittings that are closely related to the figures in the main text. 
Certain parts of the main text rely on the measurements and fitting results provided in this section.

\subsection{Interference patterns}

\begin{figure}[tbp]
    \centering
    \includegraphics{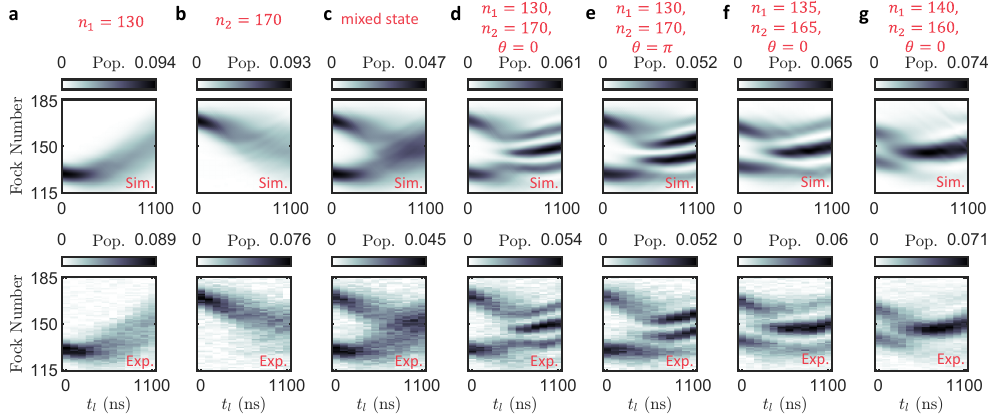}
    \caption{\textbf{2D patterns for Fock-space interferences.} 
    Tops are the simulation results and bottoms are the experimental results in each sub-figure. The color maps represent the population.
    The detuning and the strength of the single-photon pump are identical to those used in Fig.~4 of the main text.
    \textbf{a,} Diffraction of state $\ket{G_{130}}$.
    \textbf{b,} Diffraction of state $\ket{G_{170}}$.
    \textbf{c,} Propagation of the mixed state of $\ket{G_{130}}$ and $\ket{G_{170}}$, which are the averaging results of \textbf{a} and \textbf{b}.
    \textbf{d,} Interference of state $\ket{\psi_\mathrm{DG}(n_1=130,n_2=170,r_1:r_2=1,\theta=0)} $.
    \textbf{e,} Interference of state $\ket{\psi_\mathrm{DG}(n_1=130,n_2=170,r_1:r_2=1,\theta=\pi)} $.
    \textbf{f,} Interference of state $\ket{\psi_\mathrm{DG}(n_1=135,n_2=165,r_1:r_2=1,\theta=0)} $.
    \textbf{g,} Interference of state $\ket{\psi_\mathrm{DG}(n_1=140,n_2=160,r_1:r_2=1,\theta=0)} $.
    }
    \label{fig:intermore}
\end{figure}

Double-slit interference is essentially the superposition of two coherent single-slit diffraction patterns. To demonstrate the importance of coherence in this process, we conduct a control experiment with incoherent states, as shown in Figs.~\ref{fig:intermore}a-c. In these experiments, the two components of the DG state are mapped from the qubit's $\ket{\mathrm{g}}$ and $\ket{\mathrm{e}}$ states. To illustrate the evolution of the incoherent double-slit, the qubit is  initially prepared in a mixed state (i.e., 50\% in $\ket{\mathrm{g}}$ and 50\% in $\ket{\mathrm{e}}$). Aside from this initialization, all other sequences are identical to those used in Fig.~\ref{fig:allseq}d.

Figures~\ref{fig:intermore}a and \ref{fig:intermore}b show the single-slit diffraction patterns of the two components. Since a Gaussian shape remains Gaussian after a Fourier transform, no diffraction fringes are observed. Their incoherent averaging, shown in Fig.~\ref{fig:intermore}c, exhibits no interference fringes, consistent with our predictions. 
Compared to the coherent cases shown in Figs.~\ref{fig:intermore}d-e, the role of coherence is intuitively evident. The initial and final population distributions of Figs.~\ref{fig:intermore}d-e are plotted in Figs.~4b and 4c in the main text. 
 
In addition, we study how the interference fringes vary with the double-slit spacing. The interference fringes corresponding to spacings of $d=30$ and $20$ are shown in Figs.~\ref{fig:intermore}f and \ref{fig:intermore}g, respectively.
It can be observed that larger spacings result in denser interference fringes. 
Due to the finite overlap of the diffraction patterns, denser fringes allow more fringes to be visible.
In Fig.~\ref{fig:intermore}d and Fig.~4c of the main text, up to five interference fringes are displayed, which is close to the limit at the current Fock number level. 

\subsection{Fitting for obtaining the stripe spacing of the interference}

\begin{figure}[tbp]
    \centering
    \includegraphics{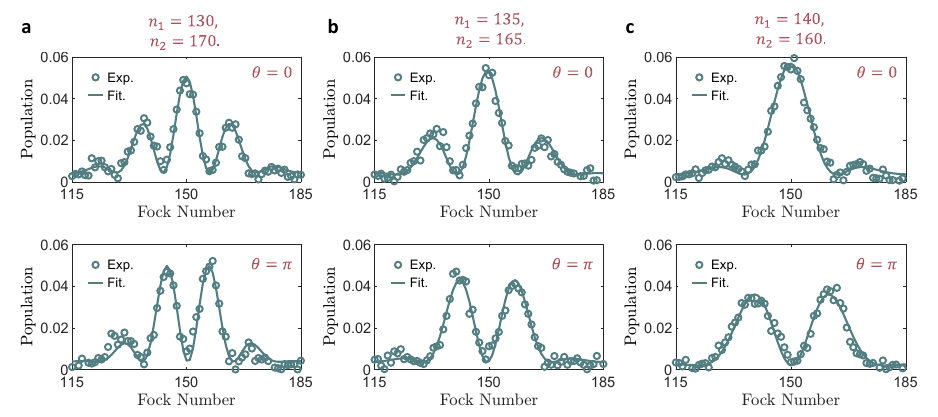}
    \caption{\textbf{Fitting for obtaining the stripe spacing of the interference.}
    We use Eq.~(\ref{eq:GsCos}) to fit the data of the interference experiment. Top row is for $\theta=0$ and bottom row is for $\theta=\pi$. The double-slit spacing $D=n_2-n_1=40$, 30, and 20 for \textbf{a}, \textbf{b}, and \textbf{c} respectively.
    }
    \label{fig:fitinter}
\end{figure}

In this section, we demonstrate how the fringe spacing in Fig.~4e of the main text is calculated. We conduct three sets of experiments for $d=40$, 30, and 20. Within each set, we measure the interference fringes for both constructive interference ($\theta=0$) and destructive interference ($\theta=\pi$), as shown in Fig.~\ref{fig:fitinter}. These data are fitted using a Gaussian-enveloped cosine function, given by 
\begin{equation} \label{eq:GsCos}
    f(n) = A e^{-\alpha_0 (n-n_0)^2} \cos\bigg[\frac{2\pi}{x}(n-n_0) + \theta \bigg], 
\end{equation}
where $n_0\approx150$ and $\theta$ are set by the experimental setup, $x$ represents the fitted fringe spacing, with the following results:
$x = 14.3~(14.0,~14.5)$ and $x=14.0~(13.6,~14.3)$ for Fig.~\ref{fig:fitinter}a top and bottom, respectively; 
$x = 18.5~(18.0,~19.0)$ and $x=19.8~(18.9,~20.7)$ for Fig.~\ref{fig:fitinter}b top and bottom respectively; 
$x = 28.2~(27.0,~29.4)$ and $x=28.4~(24.9,~31.9)$ for Fig.~\ref{fig:fitinter}c top and bottom, respectively.
We use the averaging results of the top and bottom to obtain the fringe spacing displayed in Fig.~4e (Exp.) of the main text.
We use the larger upper-limit and the smaller lower-limit from the top and bottom of each group as the error bars shown in Fig.~4e (Exp.).
Additionally, we simulate the interference patterns for 
$\theta=0$ with finer spacing variations for 
$d=14,\,16,\,18,\, ...,\,60$. 
Similarly to the analysis in Fig.~\ref{fig:fitinter}, we perform fitting for these cases, and the resulting data are presented in Fig.~4e (Sim.). 
Finally, we linearly fit the relationship between the fringe spacing and $1/d$ for the simulation data, which is plotted as an extended dotted line in Fig.~4e (Fit.), demonstrating that the line passes through the origin. 
This confirms that the fringe spacing is inversely proportional to $d$.

\subsection{Extraction of the self-Kerr coefficients by the spectrometer}\label{sec:selfKerr}

\begin{figure}[tbp]
    \centering
    \includegraphics{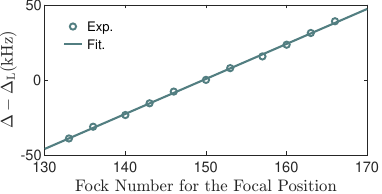}
    \caption{\textbf{Fitting for obtaining the self-Kerr coefficients by the spectrometer.} 
    For each detuning $\Delta$, we select the Fock number at the maximum of the distribution. A linear fit is then applied to these peak Fock numbers.
    }
    \label{fig:fitspec}
\end{figure}

The existence of $K_6$ will change the phase accumulation rate at large photon numbers. If $K_6$ is small, the self-Kerr can be considered approximately uniform over a certain range of photon numbers, but the accumulated phase still varies slightly with the average photon number.
The detailed derivation is as follows.

When the qubit is in its ground state and no drive is applied, Eq.~(\ref{eq:Hamitonian}) simplifies to:
\begin{equation}\begin{split}\label{eq:Hamitonian-onlycavity}
    \hat{H} & = 
     \Delta \hat{a}^\dagger\hat{a} 
     - \frac{K_\mathrm{4}}{2}\hat{a}^{\dagger2}\hat{a}^2
     - \frac{K_\mathrm{6}}{6}\hat{a}^{\dagger3}\hat{a}^3.    
\end{split}\end{equation}
After a period of time $t$, the accumulated phase distribution is:
\begin{equation}\begin{split}\label{eq:phin}
    \varphi(n)/t & =  \Delta n - \frac{K_4}{2}n(n-1) - \frac{K_6}{6}n(n-1)(n-2) \\
                 & = - \frac{K_6}{6}n^3 + (- \frac{K_4}{2} + \frac{K_6}{2})n^2
                 +(\Delta + \frac{K_4}{2} - \frac{K_6}{3})n,
\end{split}\end{equation}
where $n$ is the Fock number.
The first and second derivatives of the phase distribution are:
\begin{equation}\label{eq:phip}
    \varphi^\prime(n)/t = -\frac{K_6}{2} n^2 + (-K_4+K_6)n+(\Delta + \frac{K_4}{2} - \frac{K_6}{3}),
\end{equation}
\begin{equation}\label{eq:phipp}
    \varphi^{\prime\prime}(n)/t = -K_6 n -K_4+K_6.
\end{equation}
Equation~(\ref{eq:phipp}) gives the self-Kerr at large photon numbers.
Although $K_6$ is several orders of magnitude smaller than $K_4$, it can still lead to significant corrections at large photon numbers.

In Newton's prism experiment, the focal position is determined by the condition $\varphi^\prime=0$ for different values of $\Delta$.
From Eq.~(\ref{eq:phip}), we get
\begin{equation}
    \Delta(n_0) = \frac{K_6}{2} n_0^2 + (K_4-K_6)n_0+( -\frac{K_4}{2} + \frac{K_6}{3}),
\end{equation}
where $n_0$ represents the focal position. Although this is a quadratic function, its extremely small quadratic coefficient allows it to be approximated as a straight line within a small range. The slope of this linear approximation near $n_0$ is:
\begin{equation}
    \Delta^\prime(n_0) = K_6 n_0 +K_4-K_6.
\end{equation}
This is the same formation as that in Eq.~(\ref{eq:phipp}). Therefore we can fit the data from the Fock-space Newton's prism experiment, with the result shown in Fig.~\ref{fig:fitspec}. 
The slope of the fit is $2.33\pm0.04\,\mathrm{kHz}$, corresponding to the effective self-Kerr near $n_0=150$. Using the known value $K_4 = 2.18\pm0.09\,\mathrm{kHz}$, we have $K_6 = 1.00\pm0.87\,\mathrm{Hz}$. This approach thus provides a way to calibrate the system parameter $K_6$. 
Despite the relatively large uncertainty of $K_6$ in this experiment, we can still conclude that $K_6$ is positive.

\section{Quantum sensing with focused states from a Fock-space lens}

\begin{figure}[tbp]
    \centering
    \includegraphics{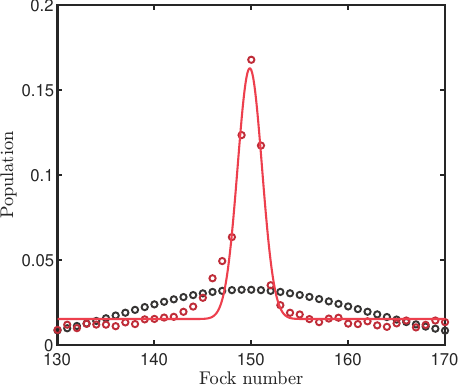}
    \caption{\textbf{Experimentally measured photon-number distribution of the 150-photon focused state.} 
     Red dots are the experimental results of the focused state.
     A Gaussian fit (red line) yields an average photon number of $\bar{n}=149.84\pm0.12$ and a standard deviation of $\sigma = 1.26\pm0.12$ photons, corresponding to a 9.7-fold compression in photon-number variance compared to the coherent state of the same average photon numbers. This compression implies a 19.7~dB quantum enhancement in small-displacement sensing over the standard quantum limit (SQL).
     For comparison, the photon-number distribution of an ideal coherent state with $\bar{n}=150$ is shown with black dots.
    }
    \label{fig:FocusedState}
\end{figure}

In the main text, we experimentally demonstrate that a Fock-space convex lens can efficiently compress the photon-number distribution of a coherent state. The resulting state at the focus exhibits an ultra-narrow photon-number distribution, making it a promising candidate for quantum-enhanced metrology. The experimentally measured photon-number distribution of the 150-photon focused state is shown in Fig.~\ref{fig:FocusedState}, revealing a Gaussian-like profile. A Gaussian fit yields an average photon number of $\bar{n}=149.84\pm0.12$ and a standard deviation of $\sigma = 1.26\pm0.12$ photons, corresponding to a 9.7-fold compression in photon-number variance compared to the coherent state of the same average photon numbers.

For small-displacement sensing, the standard quantum limit (SQL) represents the fundamental precision bound achievable with coherent states. Therefore, the achieved 9.7-fold compression implies that the displacement sensitivity of the 150-photon focused state could exceed the SQL by up to 19.7~dB. Benefiting from the scalability of our framework, this compression advantage scales with the photon number, enabling further improvement of the metrology gain. For example, the theoretical metrology gain would exceed 27~dB for a 1000-photon focused state while maintaining $\sigma < 1.3$ photons. These results underscore the potential of the Fock-space optics method in quantum sensing.


%